\documentclass[showpacs,floatfix,preprintnumbers,nofootinbib,amssymb,amsmath]{article}
\usepackage[utf8]{inputenc}
\usepackage{graphicx,bm,color,jheppub}
\usepackage{slashed,verbatim}
\usepackage{subfigure}
\usepackage[colorlinks=true,linktocpage=true,linkcolor=blue,citecolor=blue]{hyperref}

\date{}

\begin{document}
\title{Vector meson spectral function and dilepton production rate in a 
hot and dense medium within an effective QCD approach}
\author{Chowdhury Aminul Islam,}
\affiliation{Theory Division, Saha Institute of Nuclear Physics, 1/AF Bidhan Nagar, Kolkata-700064, INDIA }
\emailAdd{chowdhury.aminulislam@saha.ac.in}
\author{Sarbani Majumder,} 
\emailAdd{sarbani.majumder@saha.ac.in}
\author{Najmul Haque,}
\emailAdd{najmul.haque@saha.ac.in}
\author{Munshi G. Mustafa}
\emailAdd{munshigolam.mustafa@saha.ac.in}

\begin{abstract}
{The properties of the vector meson current-current correlation function and its spectral 
representation are investigated in details with  and without isoscalar-vector interaction 
within the 
framework of effective QCD approach, namely Nambu$\textendash$Jona-Lasinio (NJL) model 
and its Polyakov Loop extended version (PNJL), at finite temperature and finite density. 
The influence of the isoscalar-vector interaction on the vector meson correlator is 
obtained using the ring resummation known as the Random Phase  Approximation (RPA).
The spectral  as well as the correlation function in PNJL model
show that the vector meson retains its bound property  
up to a moderate value of temperature above
the phase transition.  Using the vector meson spectral function 
we, for the first time, obtained the dilepton production rate  
from a hot and dense medium within the framework of PNJL model
that takes into account the nonperturbative effect through the 
Polyakov Loop fields.
The dilepton production rate in PNJL  model is enhanced compared to 
NJL and Born rate in the deconfined phase due to the suppression of color 
degrees of freedom at moderate temperature. The presence of isoscalar-vector
interaction further enhances the dileption rate over the Born rate in the 
low mass region.
Further, we also have computed the Euclidean correlation function in vector 
channel and  the conserved density fluctuation associated with temporal 
correlation function appropriate for a hot and dense medium.
The dilepton rate and the Euclidean correlator are also compared with available 
lattice data and those quantities in PNJL model are found to agree well in certain domain.}
\end{abstract}

\maketitle
\newpage

\section{Introduction}  
Quantum Chromodynamics (QCD) being the fundamental theory of strong
interaction exhibits a very rich phase structure at extreme conditions,
{\textit {i.e.}}, high temperature and/or high density. It is now well established
that at these conditions a transition from normal hadronic matter
to a state of strongly interacting exotic quark gluon plasma (QGP)
is possible \cite{Muller:1983ed,Heinz:2000bk}. In recent
years, a tremendous effort has been devoted to the creation of  QGP in the laboratory.
The heavy-ion collider experiments presently dedicated to this search are the Relativistic Heavy
Ion Collider (RHIC) at Brookhaven National Laboratory 
(BNL)~\cite{Arsene:2004fa,Back:2004je,Adams:2005dq,Adcox:2004mh} and the Large Hadron Collider
(LHC) at the European Organization for Nuclear Research 
(CERN)~\cite{Carminati:2004fp,Alessandro:2006yt}. Also, future fixed target experiments
are planned at the Facility for Antiproton and Ion Research (FAIR) at 
the Gesellschaft f\"ur Schwerionenforschung (GSI)~\cite{Friman:2011zz}. In all these experiments heavy ions are
accelerated to relativistic speeds in order to achieve such extreme conditions required for
the creation of a short-lived QGP. Various diagnostic measurements at 
RHIC BNL~\cite{Adare:2009qk,Adler:2006yt,Adare:2006ti,Adcox:2001jp,Adcox:2002au,Adler:2003kg,
Chujo:2002bi,Adare:2006nq,Abelev:2006db} in the recent past have indicated a strong hint 
for the creation of a semi-QGP (not weakly interacting like usual QGP)  
within a first few fm/$c$ of the collisions through the manifestation of hadronic
final states. Also, new data from LHC CERN~\cite{Aamodt:2010pb,Aamodt:2010pa,
Aamodt:2010jd,Aad:2010bu,Chatrchyan:2011sx} have further 
consolidated the formation of such a state of matter.

One of the most interesting as well as 
promising subjects in the area of theoretical high energy physics is the
study of the strongly interacting many particle system under these extreme
conditions. Dynamical properties of many particle system are associated with 
the correlation function~\cite{Forster(book):1975HFBSCF,Callen:1951vq,Kubo:1957mj}. 
We know that many of the hadron properties 
are embedded in the correlation function and its spectral representation. Such 
properties in vacuum~\cite{Davidson:1995fq} are very well studied in QCD.
The presence of a stable mesonic state is understood by the
delta function like peak in the spectral function.
At low temperature  most of the hadronic models~\cite{Andronic:2012ut,Huovinen:2009yb}  work well  
but break down near the phase transition temperature, $T_c$. 
While propagating through the hot and dense 
medium~\cite{Kapusta_Gale(book):1996FTFTPA,Lebellac(book):1996TFT},
the vacuum properties of any particle get modified due to the change 
of its dispersion properties in the medium. 
The temporal correlation function is related to the response of the conserved density 
fluctuations due to the symmetry of the system. On the other hand, 
the spatial correlation function exhibits information on the masses and width.
So, in a hot and dense medium the properties of hadrons, {\textit{viz.}}, response to the fluctuations, 
masses, width, compressibility etc., will be affected. Hence, hadron properties at 
finite temperature and density are also encoded in the structure 
of its correlation function and the corresponding spectral representation, which may reflect 
the degrees of freedom  around 
the phase transition point and thus the properties of the deconfined, strongly interacting matter.
As for example, the spectral representation of the vector current-current correlation
can be indirectly accessible by high energy heavy-ion experiments as
it is related to the differential thermal cross section for
the production of lepton pairs~\cite{Kapusta_Gale(book):1996FTFTPA,Lebellac(book):1996TFT}. 
Moreover, in the limit of small frequencies, 
various transport coefficients of the hot and dense medium can be determined 
from the spatial spectral representation of the vector channel  
correlation~\cite{Kapusta_Gale(book):1996FTFTPA,Lebellac(book):1996TFT}. 
The static limit of the temporal correlation function provides the response 
to the fluctuation of the conserved quantities.
For a quasiparticle in the medium, the $\delta$-like peak is expected to be smeared
due to the thermal width, which increases with the increase in temperature.
At sufficiently high temperature and density, the contribution from the
mesonic state in the spectral function will be broad enough so that
it is not very meaningful to speak of it as a well
defined state any more~\cite{Gale:1990pn,Mustafa:1999dt,Ghosh:2009bt}.

At high temperature but zero chemical potential, the structure of the vector
meson correlation function, determination of thermal dilepton rate and various transport coefficients
have been studied using Lattice QCD (LQCD) 
framework at zero momentum~\cite{Ding:2010ga,Kaczmarek:2011ht,Francis:2011bt,Datta:2003ww,Aarts:2007wj,Aarts:2002cc,Aarts:2002tn,Amato:2013naa,Karsch:2001uw}
and also at nonzero momentum~\cite{Aarts:2005hg}, 
which is a first principle calculation that takes into account the nonperturbative effects of 
QCD. Such studies for hot and dense medium  are also performed within perturbative techniques like 
Hard Thermal Loop (HTL) 
approximation~\cite{Karsch:2000gi,Braaten:1990wp,Greiner:2010zg,Chakraborty:2003uw,Chakraborty:2001kx,
Chakraborty:2002yt,Laine:2011xm,Czerski:2008zz,Czerski:2006ct,Alberico:2006wc,Alberico:2004we,Arnold:2000dr,Arnold:2003zc,Burnier:2012ze}, 
and in dimensional reduction~\cite{Laine:2003bd,Hansson:1991kb} appropriate for weakly interacting QGP. 
Nevertheless,  at RHIC and LHC energies the 
maximum temperature reached is not very far from the phase transition temperature $T_c$ and a hot
and dense matter created in these collisions is nonperturbative in nature (semi-QGP). 
So, most of the perturbative methods may not be applicable in this temperature domain but 
these methods, however, are very reliable and accurate at very  high 
temperature~\cite{Haque:2013sja,Haque:2014rua,Haque:2013qta,Haque:2012my} for usual weakly
interacting QGP. In effective QCD model 
framework~\cite{Gale:1990pn,Mustafa:1999dt,He:2003jza,He:2002pv,He:2002ii,Kunihiro:1991qu,Oertel:2000jp}, 
several studies have also been done in this direction. 
Mesonic spectral functions and the  Euclidean  correlator in scalar, pseudoscalar 
and vector channel have been discussed in Refs.~\cite{He:2003jza,He:2002pv,He:2002ii,Kunihiro:1991qu,Oertel:2000jp} 
using Nambu$\textendash$Jona-Lasinio (NJL) model. 
The NJL model has no information related to the confinement and thus does not have any nonperturbative 
effect associated with the semi-QGP above $T_c$. However, 
the Polyakov Loop extended NJL (PNJL) model~\cite{Fukushima:2003fw,Fukushima:2003fm} contains 
nonperturbative information through Polyakov 
Loop~\cite{Pisarski:2000eq,Dumitru:2001xa,Dumitru:2002cf,Dumitru:2003hp,Gross:1980br,Gocksch:1993iy,Weiss:1980rj,Weiss:1981ev} 
that suppresses the color degrees of freedom  as the Polyakov Loop expectation value 
decreases when $T\rightarrow T_c^+$. The thermodynamic properties  
of strongly interacting matter have  been studied 
extensively within the framework of NJL and PNJL 
models~\cite{Ghosh:2006qh,Mukherjee:2006hq,Ghosh:2007wy,Ratti:2005jh,Ghosh:2014zra}.  
The properties of the mesonic
correlation functions have also been studied in scalar and pseudoscalar channel
using PNJL model in Refs.~\cite{Hansen:2006ee,Deb:2009ng}. 
We intend to study, for the first time 
in PNJL model, the properties of the vector correlation function and its spectral representation to
understand the nonperturbative effect on the spectral properties, {\it e.g.,} the dilepton production rate
and the conserved density fluctuations, in a hot and dense matter created in heavy-ion collisions.

Further, the low temperature and high density part of the phase diagram is
still less explored compared to the high temperature one. 
At finite densities the effect of chirally symmetric vector channel
interaction becomes important and it is established that within the NJL or PNJL model
this type of interaction weakens the first order transition line 
\cite{Carignano:2010ac,Kashiwa:2006rc,Sakai:2008ga,Fukushima:2008wg,Fukushima:2008is,Denke:2013jmp}
in contrary to the scalar coupling which tends to favor the appearance of 
first order phase transitions. 
It is important to mention here that the determination of
the strength of vector coupling constant is crucial under model formalism.
It cannot be fixed using vector meson mass as it is beyond
the characteristic energy cut-off of the model.
However, at the same time incorporation of vector interaction
is important if one intends to study the various spectral properties
of the system  at non-zero chemical potential~\cite{Kunihiro:1991qu,Jaikumar:2001jq}
appropriate for FAIR scenario~\cite{Friman:2011zz}.
In the present work, we study the nonperturbative
effect of Polyakov Loop, along with the presence as well as 
the absence  of the repulsive isoscalar-vector interaction,
on the spectral function, correlation function  and  various
spectral properties, ({\it e.g.,} the rate of dilepton production, fluctuation of 
conserved charges) in a hot and dense matter. The influence of this  
repulsive vector channel interaction on the correlator and its spectral 
representation has been obtained using ring resummation. 
The results are compared with NJL model
and the available lattice QCD (LQCD) data.

The paper is organized as follows: in Sec.~\ref{sc.general}
we will briefly outline some of the generalities related to the correlation
function and its spectral representation, and their relations to various physical
quantities we would like to study. In Sec.~\ref{sc.model} we briefly sketch
the effective models namely NJL and PNJL model. In Sec.~\ref{sc.ring} 
we obtain the  vector correlation function and its spectral properties
through the isoscalar-vector interaction within the ring resummation both in vacuum as well as  
in a hot and dense medium. We present our findings relevant for a hot and dense medium
in Sec.~\ref{sc.result} and finally, conclude in Sec.~\ref{sc.conclusion}.

\section{Some Generalities}
\label{sc.general}
\subsection{Correlation Function and its Spectral Representation}
In general the correlation function in coordinate space is given by
\begin{eqnarray}
{\cal G}_{AB}(t,{\vec x})\equiv 
{\cal T} ({\hat A}(t,{\vec x}){\hat B}(0,{\vec 0}))
= \int \frac{d\omega}{2\pi} \int \frac{d^3{ q}}{(2\pi)^3}
 \ \ e^{i\omega t-i{\vec q}\cdot{\vec x}} \ \ 
{\cal G}_{AB}(\omega,{ \vec q}),
\end{eqnarray}
where ${\cal T}$ is the time-ordered product of the two operators $\hat A$ and $\hat B$, 
and the four momentum $Q\equiv (\omega, {\vec q})$ with $q=|{\vec q}|$.

By taking the Fourier transformation one can obtain the momentum space
correlation function as
\begin{eqnarray}
{\cal G}_{AB}(\omega,{\vec q})
&=& \int dt \int d^3{\vec x} \ \
{\cal G}_{AB}(t,{\vec x}) 
\ \ e^{-i\omega t+i{\vec q}\cdot{\vec x}} \ \ .
\end{eqnarray}
The thermal meson current-current correlator in Euclidean time
$\tau \in $ [$0,\ \beta=1/T$] is given as \cite{Karsch:2000gi,Haque:2011iz}
\begin{eqnarray}
{\cal G}^E_{M}(\tau, {\vec x})=\langle {\cal T}(J_M(\tau, 
{\vec x})
J_M^\dagger(0, {\vec 0}))\rangle_\beta 
=T\sum_{n=-\infty}^{\infty} \int
\frac{d^3q}{(2\pi)^3} \ e^{-i(\omega_n\tau+{\vec q}
\cdot {\vec x})}\ {\cal G}^E_{M}(i\omega_n, {\vec q}),
\label{eq_curr_corr}
\end{eqnarray}
where the mesonic current is defined as $J_M={\bar {\psi}}
(\tau,{\vec x})\Gamma_M \psi(\tau,{\vec x})$,
with $\Gamma_M$ =$ 1, \gamma_5, \gamma_\mu, \gamma_\mu\gamma_5$
for scalar, pseudoscalar, vector and pseudovector channel respectively.
The momentum space correlator  ${\cal G}^E_{M}(i\omega_n, \vec q)$  
at the discrete Matsubara modes $\omega_n=2\pi nT$ can be obtained as
\begin{eqnarray}
{\cal G}^E_{M}(i\omega_n, \vec q) =
- \int_{-\infty}^{\infty} d\omega \ \ 
\frac{\sigma_{M}(\omega, \vec q)}{i\omega_n-\omega} . \label{eq.corr_discr} 
\end{eqnarray}

The spectral function $\sigma_H(\omega, \vec q)$ 
for a given mesonic channel $H$, can be obtained from
the imaginary part of the momentum space Euclidean correlator in (\ref{eq.corr_discr})
by analytic continuation as
\begin{eqnarray}
\sigma_H(\omega, \vec q)&=&\frac{1}{\pi} {\mbox{Im}}
\ {\cal G}^E_H(i\omega_n=\omega+i\epsilon,\vec q), \label{eq_spectral}
\end{eqnarray}
where $H=00,ii$ and $V$ stand for temporal, spatial and vector spectral
function, respectively. The vector spectral function is expressed in terms
of temporal and spatial components as ${\rm\sigma_V}=\sigma_{00}-\sigma_{ii}$.

Using (\ref{eq_curr_corr}) and (\ref{eq.corr_discr}) one obtains the
spectral representation of the thermal correlation function in Euclidean time
but at a fixed momentum $\vec q$ as
 \begin{equation}
{\cal G}^E_H(\tau, \vec q)=\int_0^\infty\ d\omega \
\sigma_H (\omega,  \vec q) \
\frac{\cosh[\omega(\tau-\beta/2)]}{\sinh[\omega\beta/2]}. 
\label{eq.correlation_total}
\end{equation}
Because of the difficulty in analytic continuation in LQCD
the spectral function can not be obtained directly using (\ref{eq_spectral}). 
Instead a calculation in LQCD proceeds by evaluating the Euclidean correlation 
function.  Using a probabilistic application based on maximum entropy 
method (MEM)~\cite{Nakahara:1999vy,Asakawa:2000tr,Wetzorke:2000ez}, 
(\ref{eq.correlation_total}) is then inverted  to extract the 
spectral function and  thus various spectral properties are computed in LQCD.

\subsection{Vector Spectral Function and Dilepton Rate}

 The  vector meson spectral function, $\sigma_V$, and the differential
dilepton production rate are related  as~\cite{Karsch:2000gi}
\begin{equation}
  \frac{dR}{d^4xd^4Q}  =\frac{5\alpha^2} {54\pi^2} \frac{1}{M^2} 
\ \frac{1}{e^{\omega/T}-1} \ \sigma_V(\omega, {{\vec q}}) \ , \label{eq.rel_dilep_spec}
\end{equation}
where the invariant mass of the lepton pair is $M^2=\omega^2-q^2$ and 
$\alpha$ is the fine structure constant.

\subsection{{Temporal Euclidean 
Correlation Function and Response to Conserved Density Fluctuation}}

The quark number susceptibility (QNS), $\chi_q$ measures of the response of the quark number 
density $\rho$ with 
infinitesimal change in the quark chemical  potential, $\mu+\delta\mu $ and is related to
temporal correlation function  through fluctuation dissipation theorem~\cite{Kunihiro:1991qu} as 
\begin{eqnarray}
\chi_q(T) &=&\!\! \left.\frac{\partial \rho}{\partial \mu}\right |_{\mu=0} \! \!
= \!\!\! \int \ d^4x \ \left \langle J_0(0,{\vec { x}})J_0(0,{\vec { 0}})
\right \rangle \ \nonumber \\
&=& {\lim_{\vec{ q}\rightarrow 0}}\ \beta \int \frac{d\omega}{2\pi}
\frac{-2}{1-e^{-\omega/T}} {\mbox{Im}}
{\cal G}_{00}(\omega,{{\vec q}})
 =- {\lim_{\vec{ q}\rightarrow 0}} 
{\mbox {Re}}\ {\cal G}_{00}(\omega=0, \vec q),
\label{eq4}
\end{eqnarray}
where Kramers-Kronig dispersion relation has also been used. 

The quark number conservation  implies that ${\lim_{\vec{ q}\rightarrow 0}}{\mbox{Im}}
{\cal G}_{00}(\omega,{\vec{ q}}) \propto \delta(\omega)$ and the temporal spectral function 
in (\ref{eq_spectral}) becomes
\begin{equation}
\sigma_{00}(\omega, {\vec 0})=\frac{1}{\pi} {\mbox{Im}} {\cal G}_{00}(\omega,{\vec 0})
=-\omega \delta(\omega) \chi_q(T) . \label{eq.corr_chiq}
\end{equation}
The relation of the Euclidean temporal correlation function and the response to the fluctuation
of conserved number density, $\chi_q$, can be obtained  from (\ref{eq.correlation_total}) as
\begin{equation}
 {\cal G}_{00}^E(\tau T)= -T \chi_q(T), \label{eq.eucl_chiq}
\end{equation}
which is independent of the Euclidean time, $\tau$, but depends on $T$.
\section{Effective QCD Models}
\label{sc.model}
\subsection{NJL Model}
In the present work we consider 2 flavor NJL model with vector
channel interaction.
The corresponding Lagrangian is
\cite{Kunihiro:1991qu,Klevansky:1992qe,Hatsuda:1994pi,Buballa:2003qv,Davidson:1995fq}:
\begin{eqnarray}
\mathcal{L}_{\rm{NJL}} = \bar{\psi}(i\gamma_{\mu}\partial^{\mu}-m_0)\psi
+ \frac{G_{S}}{2}[(\bar{\psi}\psi)^{2}+(\bar{\psi}i\gamma_{5}\vec{\tau}\psi)^{2}]
- \frac{G_{V}}{2}(\bar{\psi}\gamma_{\mu}\psi)^2 \label{eq_njl},
\end{eqnarray}
where, $m_0 = $diag$(m_{u},m_{d})$ with $m_{u}=m_{d}$ and $\vec{\tau}$'s
are Pauli matrices. $ G_{S}$ and $G_{V}$ are the coupling
constants of local scalar type four-quark interaction and isoscalar-vector
interaction respectively. 

 The scalar four quark interaction term leads to the formation
 of chiral condensate $\sigma=\langle \bar{\psi}\psi\rangle$.
 The condensate resulting from the additional vector coupling term
 only contains the time like component and the corresponding
 condensate is $n=\langle \bar{\psi}\gamma^0 \psi\rangle$
 \cite{Buballa:2003qv,Kashiwa:2006rc}.
 The thermodynamic potential in mean field approximation is given as:
 \begin{eqnarray}
\Omega_{\rm{NJL}}&=&\frac{G_{S}}{2}\sigma^2-\frac{G_{V}}{2}n^2-
2N_fN_c\int_{\Lambda}\frac{d^3p}{(2\pi)^3}E_p \nonumber\\
 &-&2N_fN_cT\int\frac{d^3p}{(2\pi)^3} \left[{\rm{ln}}(1+e^{-(E_p-\tilde{\mu})/T)}+{\rm{ln}}(1+e^{-(E_p+\tilde{\mu})/T)} \right].
 \end{eqnarray}
 
Here $E_p=\sqrt{{\vec p}^2+M_f^2}$ is the energy of a quark with flavor $f$
having constituent mass or the dynamical mass $M_f$ and $\Lambda$ is a
finite three momentum cut-off. The thermodynamic potential depends on the dynamical fermion
mass $M_f$ and the modified quark chemical potential $\tilde{\mu}$ 
which are related to the scalar ($\sigma$) and vector ($n$) condensates through the relations
\begin{equation}
 M_f=m_0-G_S\sigma , \label{eq.massgap}
\end{equation}
and
\begin{equation}
 \tilde{\mu}=\mu-G_V n, \label{eq_mu_tilde}
\end{equation}
respectively. 

\subsection{PNJL Model}
Let us now briefly discuss PNJL model \cite{Ghosh:2006qh,Mukherjee:2006hq,Ghosh:2007wy,
Ratti:2005jh,Fukushima:2008wg,Ghosh:2014zra}, 
where, unlike NJL model
we have a couple of more mean fields  in the form of the expectation
value of the Polyakov Loop fields~\footnote{We recall that the Polyakov 
Loop expectation value $\Phi$ acts as an order parameter~\cite{Pisarski:2000eq,Dumitru:2001xa,Dumitru:2002cf,Dumitru:2003hp,
Gross:1980br,Gocksch:1993iy,Weiss:1980rj,Weiss:1981ev} for pure $SU(N_c)$ gauge theory. Given the role of 
an order parameter, if $\Phi({\bar \Phi}) = 0$ the center symmetry $Z(N_c)$ of $SU(N_c)$ is unbroken and 
there is no ionization of $Z(N_c)$ charge, which is the confined 
phase below a certain temperature. At high temperature the $Z(N_c)$ symmetry is spontaneously broken, 
$\Phi({\bar\Phi}) \ne 0$ corresponds to a deconfined phase of gluonic plasma and there are 
$N_c$ different equilibrium states distinguished by the phase $2\pi j/N_c$ with $j=0, \cdots
(N_c-1)$. We also note that the $Z(N_c)$ symmetry is explicitly broken in presence of dynamical quark, 
yet it can be considered as an approximate symmetry and $\Phi$ can still provide useful information as 
an order parameter~\cite{Islam:2014tea}.}   
$\Phi$ and its conjugate
$\bar{\Phi}$.
The Lagrangian for the 2 flavor PNJL model with vector channel interaction
is given by,
\begin{eqnarray}
 {\mathcal L}_{\rm PNJL} &=& \bar{\psi}(i\slashed D-m_0+\gamma_0\mu)\psi +
\frac{G_S}{2}[(\bar{\psi}\psi)^2+(\bar{\psi}i\gamma_5\vec{\tau}\psi)^2]
- \frac{G_V}{2}(\bar{\psi}\gamma_{\mu}\psi)^2\nonumber\\
&-& {\mathcal U}(\Phi[A],\bar{\Phi}[A],T),
\end{eqnarray}
where $D^\mu=\partial^\mu-ig{\mathcal A}^\mu_a\lambda_a/2$, 
${\mathcal A}^\mu_a$ being the $SU(3)$ background fields and $\lambda_a$'s
are the Gell-Mann matrices. The thermodynamic potential can be
obtained as~\cite{Ghosh:2006qh,Mukherjee:2006hq,Ghosh:2007wy,Ratti:2005jh}
\begin{eqnarray}
 {\Omega}_{\textrm{PNJL}} &=& {\mathcal U}(\Phi,{\bar \Phi},T) + \frac{ G_S}{2} \sigma^2 -\frac{ G_V}{2} n^2 
 \nonumber\\
 &-&2N_fT\int \frac{d^3p}{(2\pi)^3} \ln \left[1+ 3\left(\Phi +{\bar \Phi}
 e^{-(E_p-\tilde \mu)/T} \right)e^{-(E_p-\tilde \mu)/T} + e^{-3(E_p-\tilde \mu)/T} \right ]  \nonumber \\
 &-&  2N_fT\int \frac{d^3p}{(2\pi)^3} \ln \left[1+ 3\left({\bar \Phi} + \Phi
 e^{-(E_p+\tilde \mu)/T} \right)e^{-(E_p+\tilde \mu)/T} + e^{-3(E_p+\tilde \mu)/T} \right ] \nonumber \\ 
 &-&\kappa T^4 \ln[J(\Phi,{\bar \Phi})] 
  -2N_fN_c\int_{\Lambda}\frac{d^3p}{(2\pi)^3}E_p\ .
 \label{eq.thermo_pot_pnjl}
\end{eqnarray}

The effective Polyakov Loop gauge potential is parametrized
as
\begin{equation}
 \frac{{\mathcal U}(\Phi,\bar{\Phi},T)}{T^4} = 
    -\frac{b_2(T)}{2}\Phi\bar{\Phi} -
    \frac{b_3}{6}(\Phi^3+{\bar{\Phi}}^3) +
    \frac{b_4}{4}(\bar{\Phi}\Phi)^2,
\label{eq.potential}
\end{equation}
with
\begin{equation}
 b_2(T) = a_0 + a_1\left(\frac{T_0}{T}\right) + a_2\left(\frac{T_0}{T}\right)^2 +
    a_3\left(\frac{T_0}{T}\right)^3. \nonumber\\
\end{equation}
  
\begin{table}
\begin{center}
\begin{tabular}{|cccccccc|}
 \hline
 & $a_0$ & $a_1$ & $a_2$ & $a_3$ & $b_3$ & $b_4$ & $\kappa$\\
 \hline
  & 6.75 & -1.95 & 2.625  & -7.44  & 0.75 & 7.5 & 0.1\\
 \hline
\end{tabular}
\end{center}
\caption{Parameter set used in this work for the Polyakov Loop potential and the Vander Monde term.}
\label{table_parameter}
\end{table}
Values of different coefficients $a_0,\ a_1,\ a_2,\ a_3,\ b_3$ , $b_4$ and $\kappa$
have been tabulated~\cite{Ghosh:2007wy, Hansen:2006ee} in table~\ref{table_parameter}.
The Vandermonde determinant $J(\Phi,{\bar \Phi})$ is given
as \cite{Ghosh:2007wy,Islam:2014tea}
\begin{equation}
J[\Phi, {\bar \Phi}]=\frac{27}{24\pi^2}\left[1-6\Phi {\bar \Phi}+
4(\Phi^3+{\bar \Phi}^3)-3{(\Phi {\bar \Phi})}^2\right].
\end{equation}
We also note that the mass gap in (\ref{eq.massgap}) and 
the modified chemical potential in (\ref{eq_mu_tilde}) 
will now have dependence on $\Phi$ and $\bar \Phi$ through $\sigma$ and $n$.

\section{Vector Meson correlator in Ring Resummation}
\label{sc.ring}

\begin{figure}
\center
 \includegraphics[scale=0.7]{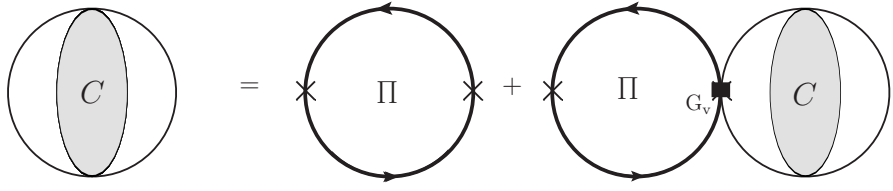}
 \caption{Vector correlator in ring resummation.}
 \label{fg.corr_resum}
\end{figure}

We intend here to consider the isoscalar-vector interaction. Generally,
from the structure of the interaction one can write the full vector 
channel correlation function  by a geometric progression of one-loop 
irreducible amplitudes \cite{Davidson:1995fq}.
In the present form of our model Lagrangian
with effective coupling $G_V$, the Dyson-Schwinger equation (DSE)
for the vector correlator $C_{\mu\nu}$ within
the ring approximation as shown in Fig.~\ref{fg.corr_resum} reads as
\begin{equation}
C_{\mu\nu} = \Pi_{\mu\nu} + G_V \Pi_{\mu\sigma} C^{\sigma}_{\nu},
\label{eq.C_munu}
\end{equation}
where $\Pi_{\mu\nu}$ is one loop vector correlator.

\subsection{Ring resummation at zero temperature and chemical potential}
The general properties of a vector correlation function at
vacuum :
\begin{eqnarray}
\Pi_{\mu\nu} (Q^2) &=& \left (g_{\mu\nu}-
\frac{Q_\mu Q_\nu}{Q^2} \right) \Pi(Q^2)
\label{eq.pi_munuk}, \\
C_{\mu\nu}(Q^2)&=& \left (g_{\mu\nu}-
\frac{Q_\mu Q_\nu}{Q^2} \right) C(Q^2),
\label{eq.c_munuk}
\end{eqnarray}
where $\Pi(Q^2)$ and $C(Q^2)$ are 
scalar quantities with $Q\equiv(q_0,{\vec q})$, is the four momentum.

Using (\ref{eq.pi_munuk}) and (\ref{eq.c_munuk}),  (\ref{eq.C_munu})  can
be reduced to a scalar DSE as
\begin{eqnarray}
  C&=& \Pi + G_V\Pi C \nonumber \\
C&=& \frac{\Pi}{1-G_V\Pi}  \label{eq8}
\end{eqnarray}

The general structure of a vector correlator in vacuum  becomes
\begin{equation}
C_{\mu\nu}= \frac{\Pi}{1-G_V\Pi} \left (g_{\mu\nu}-
\frac{Q_\mu Q_\nu}{Q^2} \right). \label{eq9}
\end{equation}
Using (\ref{eq9}) the spectral representation in vacuum can be obtained from (\ref{eq_spectral}).
The vacuum properties of vector meson can be studied using this spectral function~\cite{Davidson:1995fq}
but we are interested in those at finite temperature and density appropriate for hot and dense
medium formed in heavy-ion collisions and the vacuum, as we will see later,  is in-built therein. 
Below we briefly outline how the vector correlation function 
and its spectral properties will be modified in a hot and dense medium.

\subsection{Ring resummation at finite temperature and chemical potential}

The general structure of one-loop and resummed  vector correlation
function in the medium ($T \  {\mbox{and}} \  \mu \ne 0$) can be
decomposed~\cite{Kapusta_Gale(book):1996FTFTPA,Lebellac(book):1996TFT,Das(book):FTFT} as :
\begin{eqnarray}
\Pi_{\mu\nu} (Q^2) &=& \Pi_T(Q^2) P^T_{\mu\nu}+\Pi_L(Q^2) P^L_{\mu\nu},
\label{eq.pi_munu_T} \\
C_{\mu\nu} (Q^2) &=& C_T(Q^2) P^T_{\mu\nu}+C_L(Q^2) P^L_{\mu\nu},
\label{eq.c_munu_T}
\end{eqnarray}
where  $\Pi_{L(T)}$ and $C_{L(T)}$ are the respective scalar parts
of $\Pi_{\mu\nu}$ and $C_{\mu\nu}$.
$P_{\mu\nu}^{L(T)}$ are longitudinal (transverse) projection
operators with their well defined properties in the medium and can be chosen as \cite{Das(book):FTFT},
\begin{equation}
P_{\mu \nu}^L=\frac{Q^2}{\tilde Q^2} {\bar U_{\mu}}{\bar U_{\nu}}, ~~~~~~
P_{\mu \nu}^T=\eta_{\mu \nu}-U_\mu U_\nu -\frac{\tilde Q_\mu \tilde Q_\nu}{\tilde Q^2}.
\label{eq.p_munu}
\end{equation}
Here $U_\mu$ is the proper four velocity which in the rest frame of the heat bath
has the form $U_\mu =(1,0,0,0)$. $\tilde{Q}_\mu=(Q_\mu-\omega U_\mu)$ is the four 
momentum orthogonal to $U_\mu$ whereas
$\bar U_\mu=(U_\mu-\omega Q_\mu/Q^2)$ is orthogonal component of $U_\mu$'s with 
respect to the four momentum $Q_\mu$.
Also,  the respective scalar parts, $\Pi_{L(T)}$,
of $\Pi_{\mu\nu}$ are obtained as
\begin{equation}
\Pi_L=-\frac{Q^2}{{q}^2}\Pi_{00} ;~~~~~~~
 \Pi_T=\frac{1}{(D-2)}\left [\frac{\omega^2}{{q}^2}\Pi_{00}-\Pi_{ii}\right ],
 \label{eq.pi_munu}
\end{equation}
where $D$, is the space-time dimension of the theory.

Using (\ref{eq.pi_munu_T}) and (\ref{eq.c_munu_T}) in (\ref{eq.C_munu}) one obtains
\begin{eqnarray}
C_T(Q^2) P^T_{\mu\nu}+C_L(Q^2) P^L_{\mu\nu} &=& \left [\Pi_T
+G_V \Pi_T C_T \right] P^T_{\mu\nu}
+\left [\Pi_L+G_V \Pi_L C_L \right] P^L_{\mu\nu}.
\end{eqnarray}
Now, the comparison of coefficients on both sides leads to two scalar DSEs:  
One, for transverse mode, reads as
\begin{eqnarray}
C_T&=& \frac{\Pi_T}{1-G_V\Pi_T}, 
\label{eq.cT}
\end{eqnarray}
and the other one, for the longitudinal mode, reads as:
\begin{eqnarray}
 C_L&=& \frac{\Pi_L}{1-G_V\Pi_L} 
 \label{eq.cL}
\end{eqnarray}

Let us first write the temporal component of the resummed correlator:
 it is clear from (\ref{eq.p_munu}) that $P_{00}^T=0$. So we have
 from (\ref{eq.c_munu_T})
\begin{eqnarray}
 C_{00}&=&\frac{\Pi_L}{1-G_V \Pi_L}P_{00}^L
 = \frac{\Pi_{00}}{1+G_V\frac{Q^2}{{q}^2}\Pi_{00}}, \label{eq_c00}
\end{eqnarray}
and  the imaginary part of the temporal component of $C_{00}$ is obtained as
\begin{equation}
 {\rm Im C_{00}}=\frac{{\rm Im \Pi_{00}}}
 {\Big[ 1- G_V \Big(1-\frac{\omega^2}{{q}^2}\Big){\rm Re \Pi_{00}}\Big]^2+
 \Big[G_V (1-\frac{\omega^2}{{q}^2})
 {\rm Im \Pi_{00}}\Big ]^2} . \label{eq.C00}
\end{equation}
The spatial component of the resummed correlator ($C_{ii}$) can be written as:
\begin{equation}
 C_{ii}= C_T P_{ii}^T+C_L P_{ii}^L  =\frac{\Pi_T}{1-G_V \Pi_T}P_{ii}^T+ \frac{\Pi_L}
 {1-G_V \Pi_L}P_{ii}^L . 
\end{equation}
Using (\ref{eq.cT}) and (\ref{eq.cL}), it becomes for $D=4$
\begin{eqnarray}
 C_{ii}&=& \frac{\Pi_{ii}-\frac{\omega^2}{q^2}\Pi_{00}}{1-
 \frac{G_V}{2}(\frac{\omega^2}{ q^2}\Pi_{00}-\Pi_{ii})}
 + \frac{\frac{\omega^2}{ q^2}\Pi_{00}}
 {1+G_V(\frac{Q^2}{ q^2})\Pi_{00}} \nonumber \\
 &=& C_T' + C_L'.
\end{eqnarray}
The imaginary part of the spatial vector correlator can be obtained as
\begin{eqnarray}
 {\mbox{Im}}C_{ii}= {\mbox{Im}}C_T' + {\mbox{Im}}C_L'.
\end{eqnarray}
where
\begin{equation}
 {\rm Im}C_T'=\frac{{\rm{Im}}\Pi_{ii}-\frac{\omega^2}
 {{q}^2}{\rm{Im}}\Pi_{00}}
 {\left[1+\frac{G_V}{2}{\rm{Re}}\Pi_{ii}-\frac{G_V}{2}
 \frac{\omega^2}
 {{q}^2}{\rm{Re}}\Pi_{00}\right]^2
 +\frac{G_V^2}{4}\Big[{{\rm{Im}}\Pi_{ii}-\frac{\omega^2}
 {{q}^2}\rm{{Im}}\Pi_{00}}\Big]^2}, \label{eq.CTprime}
\end{equation}
and,
\begin{equation}
 {\rm Im}C_L'=\frac{\frac{\omega^2}{{q}^2}\rm{Im}\Pi_{00}}
 {\Big[ 1- G_V \Big(1-\frac{\omega^2}{{q}^2}\Big){\rm Re \Pi_{00}}\Big]^2+
 \Big[G_V (1-\frac{\omega^2}{{q}^2})
 {\rm Im \Pi_{00}}\Big ]^2} = \frac{\omega^2}{{q}^2}{\rm Im}C_{00}.
\label{eq.CLprime}
\end{equation}
Folowing (\ref{eq_spectral}), the resummed vector spectral function 
\begin{equation}
\sigma_V=\frac{1}{\pi}\Big[{\rm Im}C_{00}-{\rm Im}C_{ii}\Big ].
\label{eq.spectral_resum}
\end{equation}

\subsection{Conserved Density Fluctuation in Ring Approximation}
Following (\ref{eq4}) the response to the conserved density fluctuation, {\textit {i.e.}}, QNS, at finite $T$ and $\mu$ 
in ring approximation can be obtained in terms of the real part of the temporal correlation function $C_{00}$.
The real part of $C_{00}$ is  obtained from (\ref{eq_c00}) as
\begin{eqnarray}
{\rm{Re}}C_{00}(\omega,\vec q) 
& =& \frac{{\rm{Re}}\Pi_{00}(\omega, \vec q)+G_V\left (\frac{\omega^2}{ q^2}-1\right )\left[({\rm{Re}}\Pi_{00}(\omega, \vec q))^2
 +({\rm{Im}}\Pi_{00}(\omega,  \vec q))^2\right ]}
 {1+2G_V{\left (\frac{\omega^2}{ q^2}-1\right ){\rm{Re}}\Pi_{00}(\omega, \vec q)}
 +\left(G_V(\frac{\omega^2}{ q^2}-1)\right)^2
 \Big [({\rm{Re}}\Pi_{00}(\omega, \vec q))^2+({\rm{Im}}\Pi_{00}(\omega, \vec q))^2\Big]}. \nonumber \\
 \label{eq_re_c00}
\end{eqnarray}
Now the resummed QNS in ring approximation becomes
\begin{equation}
\chi^R_{q}(T,\tilde\mu)=-\lim_{\vec{q}\rightarrow 0} {\rm{Re}} C_{00}(0,\vec q) 
=\frac{\chi_q(T,\tilde \mu)}{1+G_V \chi_q(T,\tilde\mu)},
\label{eq.resum_suscp} 
\end{equation}
where we note that $\lim_{ {\vec q}\rightarrow 0}{\rm{Im}}\Pi_{00}(0, \vec q)=0$ and 
one-loop $\chi_q(T,\tilde\mu)=-\lim_{\vec{q}\rightarrow 0}{\rm{Re}}\Pi_{00}(0, \vec q)$.


Now, we note that at $G_V=0$, all the resummed quantities in ring approximation become equivalent to those of one-loop. To compute all
resummed quantities, we just need to compute one-loop vector self-energy within the effective models considered here.

\subsection{Vector correlation function in one-loop}
\label{sc.resum_corr}

\begin{figure} [!htb]
\center
\includegraphics[scale=0.4]{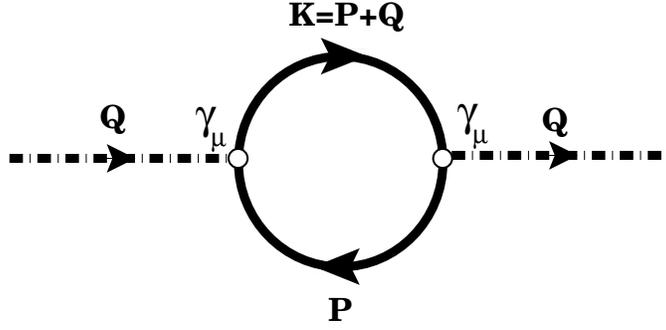}
\caption{Vector channel correlator at one-loop.}
\label{fg.corr_oneloop}
\end{figure}

The current-current correlator in vector channel
at one-loop level can be written as
\begin{equation}
 \Pi_{\mu\nu}(Q)= \int\frac{d^4 P}{(2\pi)^4} {\rm{Tr}}_{D,c}
 \left[\gamma_\mu S(P+Q) \gamma_\nu S(P)\right],
 \label{eq_pi_munu}
\end{equation}
where ${\textrm {Tr}}_{D,c}$ is trace over Dirac and color indices, respectively.
We would like to compute this in 
effective models, {\textit {viz.}}, in NJL and PNJL model. 

The NJL quark propagator in Hartree approximation is given as
\begin{equation}
{S_{\textrm{NJL}}}(L)= \left[\slashed l -m_0 +\gamma_0\tilde\mu+G_S\sigma\right ]^{-1}
= \left[\slashed l -M_f +\gamma_0{\tilde\mu}\right ]^{-1},
\label{eq.mod_prop_inv}
\end{equation}
whereas for PNJL it reads as
\begin{equation}
S_{\textrm{PNJL}}(L)=\left[\slashed l -M_f +\gamma_0{\tilde \mu} -i\gamma_0{\cal A}_4\right]^{-1} 
\label{eq.mod_prop_pnjl_inv}
\end{equation}
where the four momentum, $L\equiv(l_0,{\vec l}\hspace{0.05cm})$. The gap equation for constituent quark mass $M_f$ 
and the modified quark chemical 
potential $\tilde \mu$ due to vector coupling $G_V$ are given, respectively, in (\ref{eq.massgap}) 
and (\ref{eq_mu_tilde}). Now in contrast to NJL one, 
the presence of background temporal gauge field ${\cal A}_4$ will make a connection 
to the Polyakov Loop field $\Phi$ 
\cite{Pisarski:2000eq,Dumitru:2001xa,Dumitru:2002cf,Dumitru:2003hp,Gross:1980br,Gocksch:1993iy,Weiss:1980rj,Weiss:1981ev}. 
While performing the frequency sum and color trace in (\ref{eq_pi_munu}), the thermal distribution
function in PNJL case will be different from that of  NJL one.

For convenience, we will calculate the one-loop vector correlation in NJL model.  
This NJL correlation function, as discussed, can easily be generalized to PNJL one by replacing 
the thermal distribution function~\footnote{For details we refer to Refs.~\cite{Hansen:2006ee,Ghosh:2014zra}.} as
\begin{equation}
 f(E_p \pm \tilde\mu)=\frac{\Phi e^{-\beta(E_p \pm \tilde \mu)}
 +2\bar{\Phi}e^{-2\beta(E_p \pm \tilde \mu)}+e^{-3\beta(E_p \pm \tilde \mu)}}
 {1+3\Phi e^{-\beta(E_p \pm \tilde\mu)}+3\bar{\Phi}e^{-2\beta(E_p \pm \tilde\mu)}
 +e^{-3\beta(E_p \pm \tilde\mu)}}, \label{eq_dist_pnjl}
\end{equation}
which at $\Phi ({\bar \Phi})=1$ reduces to the thermal distribution for NJL or free as the cases may be. 
On the other hand for $\Phi ({\bar \Phi})=0$, the effect of confinement
is clearly evident in which three quarks are stacked in a same momentum and color state~\cite{Islam:2014tea}. 

\subsubsection{Temporal Part}
The time-time component of the vector correlator in (\ref{eq_pi_munu}) reads as
\begin{equation}
 \Pi_{00}(Q)=\int{\frac{d^4P}{(2\pi)^4}{\mathrm {Tr}}[\gamma_0 S(K)\gamma_0 S(P)]},
\label{eq.pi00}
\end{equation}
where $K=P+Q$. After some mathematical simplifications we are left with
\begin{eqnarray}
 \Pi_{00}(\omega,\vec q) &=& N_c N_f\int\frac{d^3p}{(2\pi)^3}\frac{1}{E_{p}E_{k}}\Bigg\{
\frac{E_{p}E_{k}+M_f^2+\vec{p}\cdot\vec{k}}{\omega+E_{p}-E_{k}} \nonumber\\
&& \times \left[f(E_{p}-\tilde \mu)+f(E_{p}+\tilde \mu)
-f(E_{k}-\tilde\mu)-
f(E_{k}+\tilde \mu)\right] \nonumber \\
&& + \left(E_{p}E_{k}-M_f^2-\vec{p}\cdot\vec{k}\right)
\left[\frac{1}{\omega-E_{p}-E_{k}}-
\frac{1}{\omega+E_{p}+E_{k}}\right] \nonumber\\
&& \times \left[1-f(E_{p}+\tilde\mu)
-f(E_{k}-\tilde\mu)\right]
\Bigg\}. \label{eq.total_pi00} 
\end{eqnarray}

The real and imaginary parts of the temporal vector correlator are, respectively,  obtained as
\begin{eqnarray}
\textrm{Re} \Pi_{00}(\omega,\vec q) &=&  {\large {\textrm  P}}\Bigg [\ N_f N_c \int\frac{d^3p}{(2\pi)^3}\frac{1}{E_pE_k}
\Bigg \{\frac{E_p E_k+M_f^2+\vec{p}\cdot\vec{k}}{\omega +E_p -E_k} \nonumber\\
&& \times \left[f(E_{p}-\tilde\mu)+f(E_{p}+\tilde\mu)-f(E_{k}-\tilde\mu)-f(E_{k}+\tilde\mu)\right] \nonumber\\
&& +(E_p E_k-M_f^2-\vec{p}\cdot\vec{k})\left(\frac{1}{\omega-E_p-E_k}-\frac{1}{\omega+E_p+E_k}\right) \nonumber\\
&& \times \left[1-f(E_p+\tilde\mu)-f(E_k-\tilde\mu)\right]\Bigg\}\Bigg ],
\label{eq.repi00}
\end{eqnarray}
as {\large P} stands for principal value,
and
\begin{eqnarray}
 {\textrm {Im}} \Pi_{00}(\omega,\vec q) &=& \lim_{\eta\rightarrow0}\ \frac{1}{2i}
\Big [\Pi_{00}(\omega\rightarrow \omega +i\eta,q)
-\Pi_{00}(\omega\rightarrow \omega -i\eta,q)\Big ] \nonumber \\
&=& -\pi N_f N_c \int\frac{d^3p}{(2\pi)^3} 
\frac{1}{E_pE_k}
\Bigg\{ 
(E_p E_k+M_f^2+\vec{p}\cdot\vec{k}) \nonumber \\
&& \times \left[f(E_{p}-\tilde\mu)+
f(E_{p}+\tilde\mu)-f(E_{k}-\tilde \mu)-f(E_{k}+\tilde\mu)\right]
\times \delta(\omega+E_p-E_k) \nonumber \\
&& + (E_p E_k-M_f^2-\vec{p}\cdot\vec{k})
\left[1-f(E_p+\tilde\mu)-
f(E_k-\tilde \mu)\right] \nonumber \\
&& \times [\delta(\omega-E_p-E_k)]\Bigg\}.
\label{eq.im_pi}
\end{eqnarray}
It is now worthwhile 
to check some known results in the limit  
$\vec q \rightarrow 0$ and $\tilde\mu=0$,  (\ref{eq.im_pi}) can be written as: 
\begin{eqnarray}
\textrm{Im} \Pi_{00}(\omega)&=& -\pi N_f N_c \int\frac{d^3p}
{(2\pi)^3}\frac{1}{E_p^2}
\left (3E_p^2-3M_f^2-p^2 \right) \left(2f'(E_p)\right)
\left(-\omega \delta\left(\omega \right)\right),
 \label{eq.impi_00_q0_2}
\end{eqnarray}
and which, in the limit $M_f=m_0-G_S\sigma=0$, further becomes
\begin{eqnarray}
\textrm{Im} \Pi_{00}(\omega)
&=& -2\pi T^2\omega \delta(\omega).
\label{eq.impi_00_q0_m0}
\end{eqnarray}

The vacuum part in (\ref{eq.repi00}) is now separated as 
\begin{eqnarray}
{\textrm {Re}}\Pi_{00}^{\textrm {vac}}(\omega,\vec q)\!\! &=& \!\!
\frac{N_f N_c}{4\pi ^2} \!\! \int _0^{\Lambda } \!\! p\ dp
\frac{1}{2 E_p q}\left [ 4 p q+6 E_p X_- -6 E_p X_+ 
- Y_- \ln\left | \frac{E_p+ X_- -\omega } {E_p+ X_+ -\omega }\right| \right. \nonumber \\
&&\left. + Y_+  \ln\frac{E_p+ X_+ +\omega } {E_p + X_- +\omega }\right] , \label{eq.repivac00} \\
{\textrm{with}} \, \, \, \,  Y_\pm &=& (4 E_p^2 \pm 4 E_p \omega +M^2),  \, \, \, \,
X_\pm =\sqrt{E_p^2\pm 2 p q+q^2} \ \ \ {\textrm{and}} \ \ \ \  M^2=\omega^2-q^2.
\end{eqnarray}
We note that the  ultraviolet divergence in the vacuum part is 
regulated by using  a finite three momentum cut-off $\Lambda$. The corresponding matter 
part of (\ref{eq.repi00}) is obtained as
\begin{eqnarray}
{\textrm {Re}} \Pi_{00}^{\textrm {mat}}(\omega,\vec q)&=&\frac{N_f N_c}{2\pi ^2}\int _0^{\infty }p \ dp
\big[f(E_p-\tilde\mu)+f(E_p+\tilde\mu) \big]
\left[\frac{\omega }{q}\ln\left |\frac{M^2-4\left(pq+\omega E_p\right)^2}
{M^2-4\left(pq-\omega E_p \right)^2}\right | \right. \nonumber\\
&&\left. 
-\left(\frac{4E_p^2 + M^2}{4qE_p}\right)
\ln\left | \frac{\left(M^2-2pq\right)^2-4\omega ^2 E_p^2}{\left(M^2+2pq\right)^2-4\omega^2 E_p^2}\right |
-\frac{2p}{E_p}\right] . \label{eq.repimat00}
\end{eqnarray}

The imaginary part in (\ref{eq.im_pi}) can be simplified as
\begin{eqnarray}
{\textrm {Im}} \Pi_{00}(\omega,\vec q)&=& \frac{ N_f N_c}{4\pi }\!\!\! \int _{p_-}^{p_+}
\!\!\!  p \ dp \ \frac{4\omega E_p -4E_p^2-M^2}{2E_p q} \ \big [f(E_p-\tilde\mu)+ f(E_p+ \tilde \mu) - 1 \big ]
\label{eq.impimat00}
\end{eqnarray}
where the vacuum part does not require any momentum cut-off as 
the energy conserving $\delta$-function 
ensures the finiteness of the limits:
\begin{equation}
p_\pm = {\frac{\omega }{2}\sqrt{1-\frac{4M_f^2}{M^2}}\pm\frac{q}{2}}, \label{eq.limits}
\end{equation}
with a threshold restricted by a step function, $\Theta(M^2-4M_f^2)$.

\subsubsection{Spatial Part}
The space-space component of the vector correlator in (\ref{eq_pi_munu}) reads as
\begin{equation}
 \Pi_{ii}(Q)=\int{\frac{d^4P}{(2\pi)^4}{\mathrm {Tr}}
 [\gamma_i S(K)\gamma_i S(P)]},
\label{eq_pi_ii}
\end{equation}
which can be simplified to
\begin{eqnarray}
 \Pi_{ii}(\omega,\vec q) &=& N_c N_f\int\frac{d^3p}{(2\pi)^3}\frac{1}
 {E_{p}E_{k}}\Bigg\{
 \frac{3E_{p}E_{k}-3M_f^2-\vec{p}\cdot\vec{k}}{\omega-E_{p}+E_{k}}
 \nonumber \\
 && \times \left[f(E_k+\tilde\mu)+f(E_k-\tilde\mu) - f(E_p+\tilde\mu)-f(E_p-\tilde\mu)\right]
 \nonumber \\
&&+ \left (3E_{p}E_{k}+3M_f^2+\vec{p}\cdot\vec{k}\right ) 
 \left [\frac{1}{\omega-E_p-E_k} - \frac{1}{\omega+E_p+E_k}\right ]
 \nonumber \\
&&\times  \left [1-f(E_k+\tilde\mu)-f(E_p-\tilde\mu) \right ]
\Bigg\}. \label{eq.total_pi_ii} 
\end{eqnarray}

In the  similar way as before the imaginary part can be obtained 
\begin{eqnarray}
 {\textrm {Im}} \Pi_{ii}(\omega,\vec q) &=& -\pi N_f N_c
 \int\frac{d^3p}{(2\pi)^3} 
 \frac{1}{E_pE_k}
\Bigg\{(3E_{p}E_{k}-3M_f^2-\vec{p}\cdot\vec{k}) \nonumber\\
&& \times \left[f(E_k+\tilde\mu)+f(E_k-\tilde\mu) - f(E_p+\tilde\mu)-f(E_p-\tilde\mu)
\right]\delta(\omega+E_p-E_k) \nonumber \\
&& + \left (3E_{p}E_{k}+3M_f^2+\vec{p}\cdot\vec{k}\right )
\left [1-f(E_k+\tilde\mu)-f(E_p-\tilde\mu) \right ] \nonumber\\
&& \times \delta(\omega-E_p-E_k)\Bigg\},
\label{eq.impi_ii}
\end{eqnarray}
whereas the real part can be obtained as
\begin{eqnarray}
\textrm{Re} \Pi_{ii}(\omega,\vec q) &=& {\large {\textrm  P}} \Bigg[ N_f N_c \int
\frac{d^3p}{(2\pi)^3}\frac{1}{E_pE_k}
\Bigg \{\frac{3E_p E_k-3M_f^2-\vec{p}\cdot\vec{k}}
{\omega -E_p +E_k} \nonumber\\
&& \times \left[f(E_{k}-\tilde\mu)+f(E_{k}+\tilde\mu)
-f(E_{p}-\tilde\mu)-f(E_{p}+\tilde \mu)\right] \nonumber\\
&& + (3E_p E_k+3M_f^2+\vec{p}\cdot\vec{k})
\left(\frac{1}{\omega-E_p-E_k}
-\frac{1}{\omega+E_p+E_k}\right) \nonumber\\
&& \times \left[1-f(E_k+\tilde \mu)-f(E_p-\tilde \mu)\right]\Bigg\}\Bigg].
\label{eq.repi_ii}
\end{eqnarray}

At this point we also check the known results in the limit 
$\vec q \rightarrow 0$ and $\tilde\mu=0$,  (\ref{eq.impi_ii}) can be written as: 
\begin{eqnarray}
\textrm{Im} \Pi_{ii}(\omega)&=& -\pi N_f N_c \int\frac{d^3p}
{(2\pi)^3}\frac{1}{E_p^2}
\left (3E_p^2-3M_f^2-p^2 \right) \left(2f'(E_p)\right)
\left(-\omega \delta\left(\omega \right)\right) \nonumber \\
&& - \frac{3}{2\pi\omega}\sqrt{\omega^2-4M_f^2}
\left(\omega^2+2M_f^2\right)\tanh
\left(\frac{\omega}{4T}\right)\Theta(\omega-2M_f),
 \label{eq.impi_ii_q0_2}
\end{eqnarray}
and when $M_f=m_0-G_S\sigma=0$, it becomes
\begin{eqnarray}
\textrm{Im} \Pi_{ii}(\omega)
&=& -2\pi T^2\omega \delta(\omega)-
{\frac{3}{2\pi}}\omega^2 \tanh
\left(\frac{\omega}{4T}\right).
\label{eq.impi_ii_q0_m0}
\end{eqnarray}
As seen both massive and massless cases show a sharp peak due to the delta function at $\omega \rightarrow 0$, which 
leads to pinch singularity for calculation of transport coefficients.

Now, the vacuum part in (\ref{eq.repi_ii}) is simplified as
\begin{eqnarray}
{\textrm {Re}} \Pi^{\textrm{vac}}_{ii}(\omega,\vec q)\!\!\!&=&\!\!\! \frac{N_f N_c}{4\pi ^2}
\!\! \int _0^{\Lambda }\!\!\!\! p\ dp \
\frac{1}{2 E_p q}\left[-4 p q+10 E_p (X_- - X_+) - Z_-
\ln\left |\frac{E_p+X_--\omega }{E_p+X_+ -\omega }\right | \nonumber \right. \nonumber \\ 
&& \left. + Z_+ 
\ln \frac {E_p+X_+ +\omega } {E_p+X_-+\omega }\right] , \label{eq.repivacii}
\end{eqnarray}
where $Z_\pm=4p^2 \pm 4 E_p \omega -M^2$. The corresponding matter part is obtained as
\begin{eqnarray}
{\textrm {Re}} \Pi_{ii}^{\textrm{mat}}(\omega,\vec q)&=&\frac{N_f N_c}{4\pi ^2}\int _0^{\infty }p\ dp \ 
\big[f(E_p-\tilde\mu)+f(E_p+\tilde\mu)\big]
\left[2\frac{\omega}{q}\ln\left |\frac{M^2-4(pq+E_p \omega)^2}{M^2-4(pq-E_p\omega)^2}\right| \right. \nonumber \\
&& \left. +\left(\frac{M^2-4p^2}{2qE_p}\right)
\ln\left |\frac{\left(M^2-2pq\right)^2-4\omega^2E_p^2}{\left(M^2+2pq\right)^2-4\omega^2 E_p^2}\right |+4\frac{p}{E_p}\right] .
\label{eq.repimatii}
\end{eqnarray}
Finally, the imaginary part in (\ref{eq.impi_ii}) is simplified as
\begin{eqnarray}
{\textrm {Im}} \Pi_{ii}(\omega,\vec q)&=&\frac{ N_f N_c}{4\pi } \int _{p_-}^{p_+}p \ dp \
\frac{4\omega E_p-4p^2+M^2}{2E_p q}\ \left (f(E_p-\tilde\mu)+ f(E_p+ \tilde\mu) - 1 \right )
\label{eq.impimatii}
\end{eqnarray}
where the vacuum part does not need any finite momentum cut-off as stated above.

\section{Results}
\label{sc.result}
\subsection{Gap Equations and Mean Fields}
\begin{figure} [!htb]
\subfigure[]
{\includegraphics[scale=0.8]{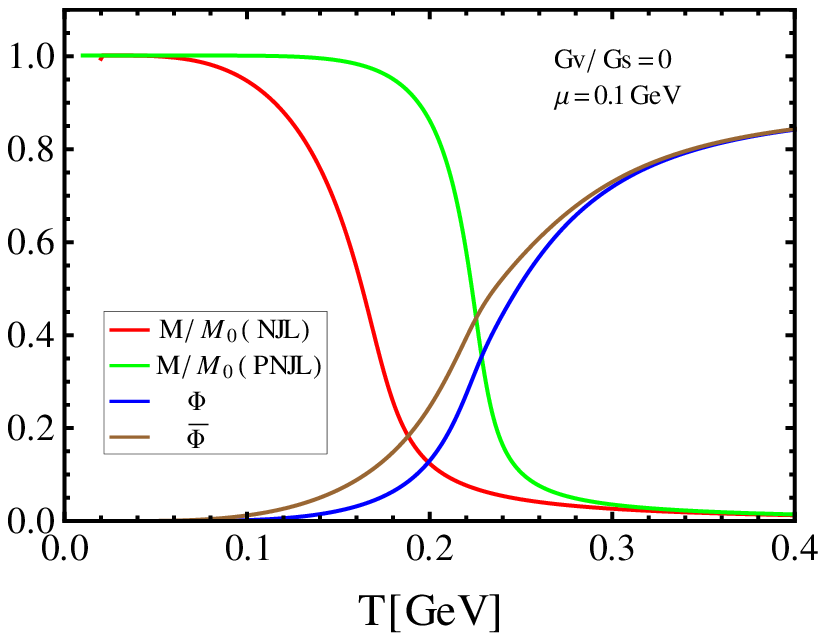}
\label{fg.massfields_gv0}}
\subfigure[]
 {\includegraphics[scale=0.8]{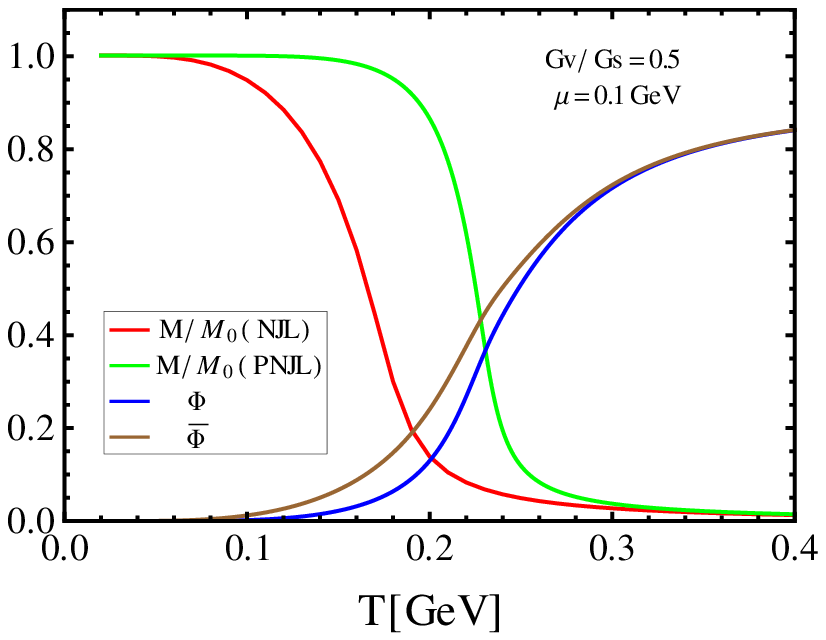}
 \label{fg.massfields_gv05}}
 \caption{Variation of scaled constituent quark mass with zero temperature quark mass 
 and the Polyakov Loop fields ($\Phi$ and $\bar \Phi$) with temperature 
 at chemical potential $\mu=100$ MeV for (a) $G_V/G_S=0$ 
 and (b) $G_V/G_S=0.5$.}
\label{fg.fields&mass_vs_T}
\end{figure}

\begin{figure} [!htb]
\subfigure[]
{\includegraphics[scale=0.8]{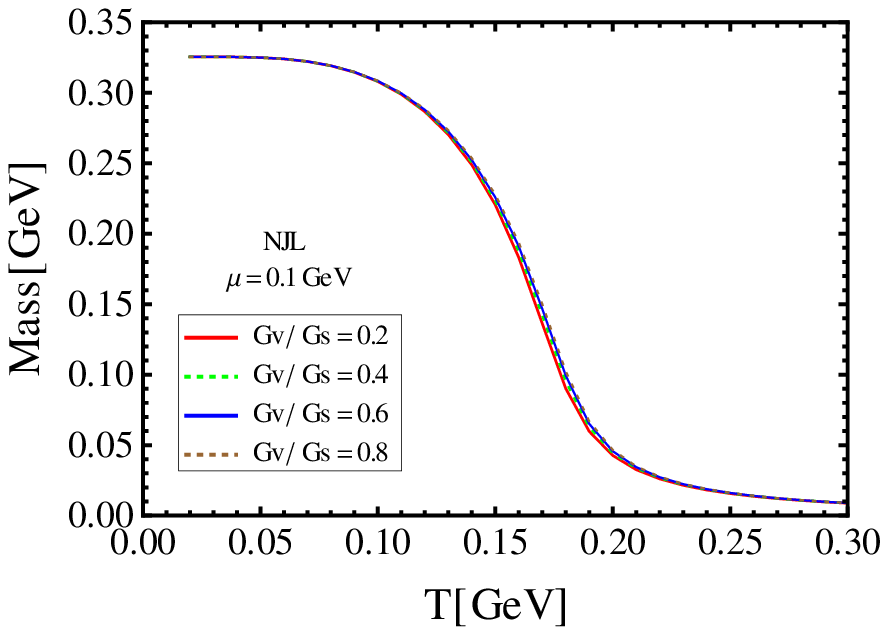}
\label{fg.mass_njl}}
\subfigure[]
 {\includegraphics[scale=0.8]{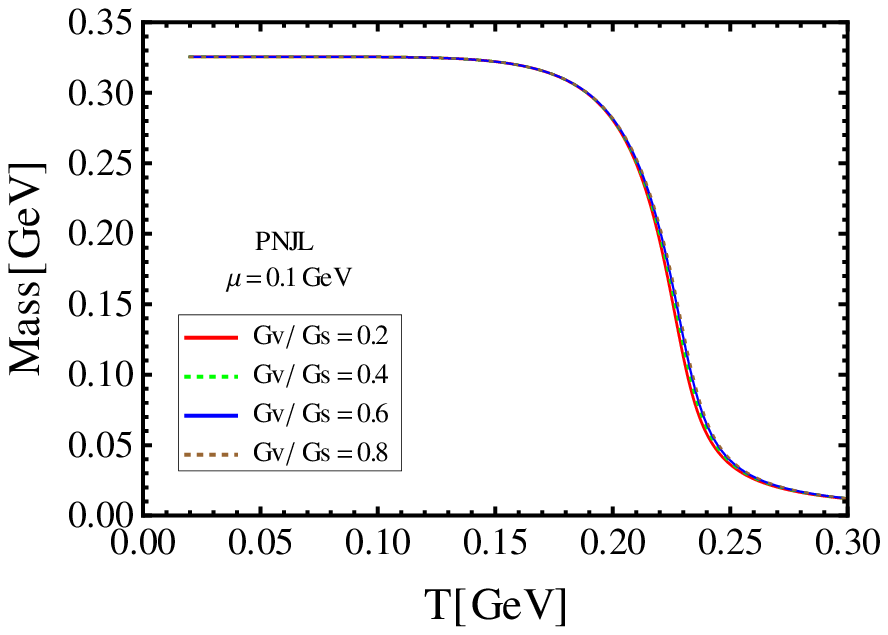}
 \label{fg.mass_pnjl}}
 \caption{Variation of constituent quark mass with temperature in  (a) NJL  and 
(b) PNJL  model for chemical potential $\mu=100$ MeV and a set of
 values for vector coupling $G_V$.}
\label{fg.mass_vs_T}
\end{figure}

The thermodynamic potential $\Omega$ for both NJL and PNJL model is extremized 
with respect to the mean fields X, {\textit{i.e.}},
\begin{equation}
 \frac{\partial \Omega}{\partial X}=0
\end{equation}
where, X stands for $\sigma$ and $n$ for NJL model and 
$\Phi, \bar{\Phi},\sigma$ and $n$ for PNJL model.
The value of the parameters, $G_S=10.08 \ {\textrm{GeV}}^{-2}$ and
$\Lambda=0.651$ GeV were taken from literature \cite{Ratti:2005jh}
and $m_0=0.005$ GeV.
However, the value of $G_V$ is difficult to fix within
the model formalism,
since this quantity should be fixed using the $\rho$ meson
mass which, in general, happens to be higher than the
maximum energy scale $\Lambda$ of the model. 
So, we consider the vector coupling constant $G_V$ as a free parameter and
different choices are considered as ${G_V}=x \times {G_S}$, where $x$ is chosen
from $0$ to $1$ appropriately.

\begin{figure} [!htb]
\subfigure[]
{\includegraphics[scale=0.8]{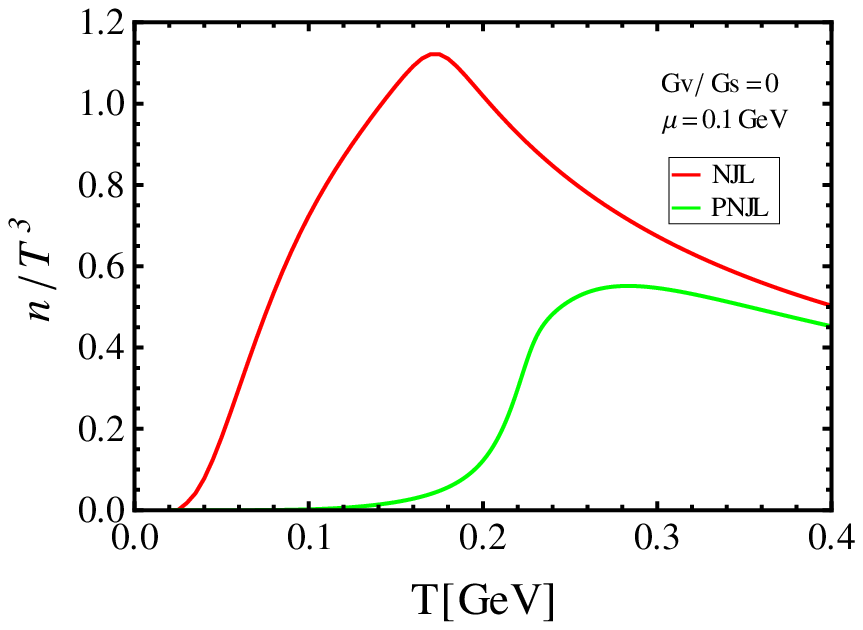}
\label{fg.density_gv0}}
\subfigure[]
 {\includegraphics[scale=0.8]{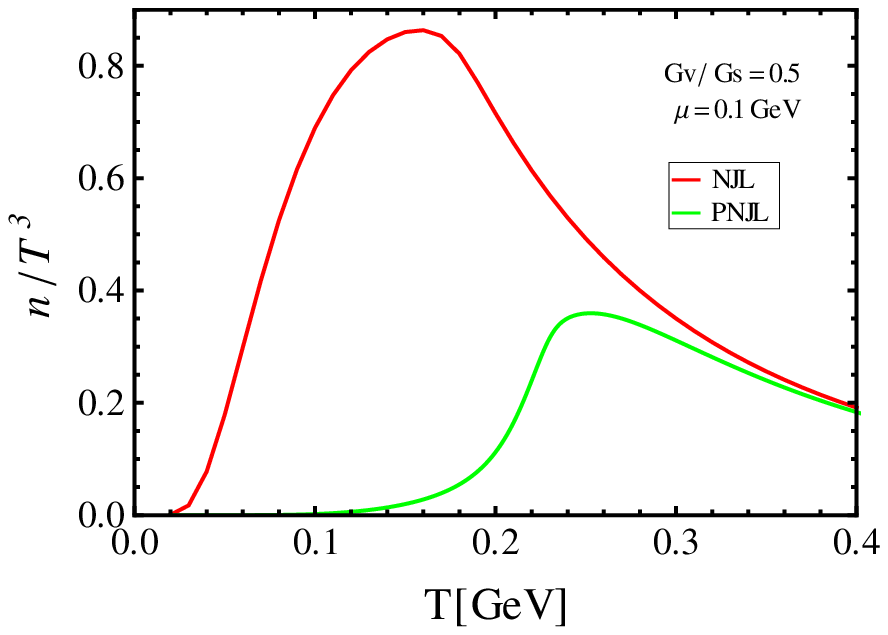}
 \label{fg.density_gv05}}
 \caption{Comparison of the scaled quark number density with $T^3$ 
 as a function of  $T$ in  NJL and PNJL model with chemical potential $\mu=100$ MeV
 for (a) $G_V/G_S=0$ and (b) $G_V/G_S=0.5$ .}
\label{fg.number_density}
\end{figure}

In Fig.~\ref{fg.fields&mass_vs_T} a comparison between the scaled quark mass with its zero temperature
value 
($M_f(T)/M_f(0)$) in NJL and PNJL model is displayed as a function of temperature $T$ for two values 
of $G_V (=0 \ {\textrm{and}} \ 0.5G_S)$ with $\mu=100$MeV. It also contains a variation of the Polyakov loop fields
($\Phi(T)$ and $\bar \Phi(T)$) with $T$. 
In both models the scaled quark mass decreases with increase in $T$ and approaches 
the chiral limit at very high $T$. However, in the temperature range $60\le T({\textrm{MeV}})\le 300$,
the variation of the scaled quark mass is slower in PNJL than that of NJL model. This slow variation is
due to the presence of the confinement effect in it through the Polyakov Loop fields ($\Phi(T)$ and $\bar \Phi(T)$),
as can be seen that the  Polyakov Loop field ($\Phi(T)$) and its conjugate ($\bar \Phi(T)$) increase from zero 
in confined phase and approaches unity (free state) at high temperature. Now, we note that $\Phi={\bar \Phi}$ 
at $\mu=0$ as there is equal number of quarks and antiquarks.  However, because of non-zero chemical potential 
there is an asymmetry in 
quark and antiquark numbers, which leads to an asymmetry in the Polyakov Loop fields $\Phi$ and $\bar \Phi$. 
This asymmetry disappears for $T> 300$ MeV, which is much greater than $\mu$. 
We also note that the fields depend weakly with the variation of the vector coupling $G_V$ akin to that 
of mass as displayed in Fig.~\ref{fg.mass_vs_T} for $\mu=100$ MeV and $G_V/G_S=0 \ {\textrm {to}} \ 0.8$.

In Fig.~\ref{fg.number_density} the  number density scaled with $T^3$ for both NJL and PNJL model is 
displayed as a function of $T$ with  $\mu= 100$ MeV for two values of $G_V$. 
 At very high temperature, as seen in Fig.~\ref{fg.fields&mass_vs_T}, 
the Polyakov Loop fields $\Phi({\bar \Phi})\rightarrow 1$ and masses in both models
become same. So, PNJL model becomes equivalent to NJL model because the thermal distribution 
function becomes equal as can be seen from (\ref{eq_dist_pnjl}) and thus the number density. 
On the other hand, for temperature $T< 400$ MeV and a given $\mu$, the PNJL number density 
is found to be suppressed than that of NJL case as the thermal distribution function in 
(\ref{eq_dist_pnjl}) is suppressed. This is due to the combination of two complementary effects:
(i) the nonperturbative effect through $\Phi$ and ${\bar \Phi}$ 
and (ii) the slower variation of mass in PNJL model, which are clearly evident from 
Fig.~\ref{fg.fields&mass_vs_T}. It is also obvious from
Fig.~\ref{fg.density_gv0} and Fig.~\ref{fg.density_gv05} that 
the presence of vector interaction $G_V$ reduces the number density for both NJL and PNJL model, 
which could be understood due to the reduction of ${\tilde \mu}$ as given in (\ref{eq_mu_tilde}).

\subsection{Vector spectral function and dilepton rate}

 The vector spectral function is proportional to the imaginary part of the vector correlation
function as defined in (\ref{eq_spectral}) or (\ref{eq.spectral_resum}). This imaginary part is restricted by, 
the energy conservation, $\omega = E_p+E_k$, as can be seen from (\ref{eq.im_pi}) and 
(\ref{eq.impi_ii}). This equivalently leads to a threshold, $M^2\ge 4M_f^2$, which can also be found
from (\ref{eq.limits}). Now for a given $G_V$ and  $T$, the resummed spectral 
function in (\ref{eq.spectral_resum}) picks up continuous contribution  
above the threshold, $M^2> 4M^2_f$, which provides a finite width to a vector meson that decays 
into a pair of leptons.  However, below the threshold $M^2 < 4M^2_f$, the continuous contribution of
the spectral function in (\ref{eq.spectral_resum}) becomes zero and the decay to dileptons are forbidden. 
But if one analyses it below the threshold, one can find bound state contributions in
spectral functions.
When imaginary part  approaches zero, the spectral function in (\ref{eq.spectral_resum}) becomes 
discrete and can be written as:
\begin{eqnarray}
 \sigma_V(\omega,\vec q) &=\atop {M<2M_f} & \frac{1}{\pi} \Big [\delta\big (F_1(\omega,\vec q)\big )\Big], \nonumber \\
\textrm{where} \, \,  \, \, \, \, \, \, \,   
 F_1(\omega, \vec q) &=& {1+\frac{G_V}{2}{\rm{Re}}\Pi_{ii}-\frac{G_V}{2} 
 \frac{\omega^2}
 {{q}^2}{\rm{Re}}\Pi_{00}}=0,
\label{eq.delta_spect}
\end{eqnarray}
where only the dominant contribution of (\ref{eq.CTprime}) is considered.
Using the properties of $\delta$-function,
one can write
\begin{eqnarray}
 \sigma_V(\omega, \vec q) & = \atop {M<2M_f} & \frac{1}{\pi} \frac{\delta(\omega-\omega_0)}
 {\left | dF_1(\omega,\vec q)/d\omega\right |_{\omega=\omega_0}},   \label{eq.discrete_spect}
\end{eqnarray}
which corresponds to a sharp $\delta$-function peak at $\omega=\omega_0$. However, we are interested here 
in continuous contribution $M> 2M_f$, which are discussed below.

\begin{figure} [!htb]
\subfigure[]
{\includegraphics[scale=0.8]{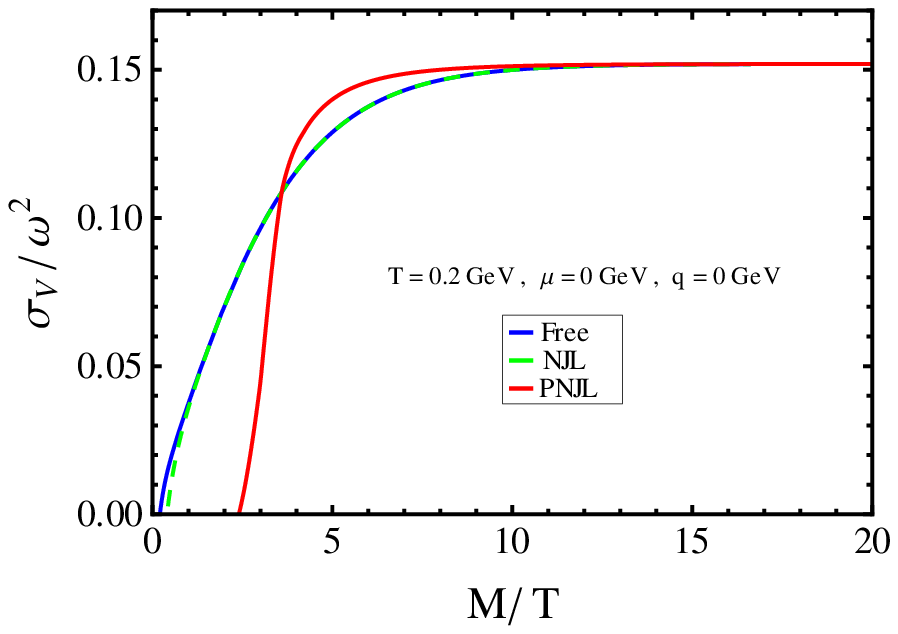}
\label{fg.free_mu0_t2}}
\subfigure[]
{\includegraphics[scale=0.8]{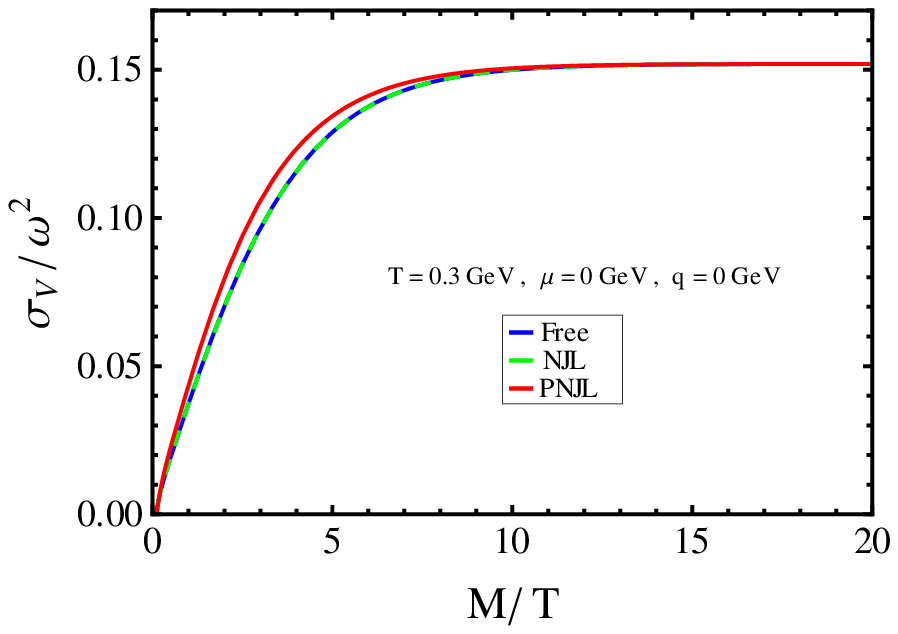}
 \label{fg.free_mu0_t3}}
\caption{Scaled vector spectral function $\sigma_V/\omega^2$ as a function of scaled 
invariant mass, $M/T$, in NJL and PNJL model with external momentum 
$q=0$, quark chemical potential $\mu=0$ and $G_V/G_S=0$ for (a) $T=200$ MeV and (b) $T=300$ MeV}
\label{fg.spect_gv0_q0_mu0}
\end{figure}

\begin{figure} [!htb]
\subfigure[]
{\includegraphics[scale=0.8]{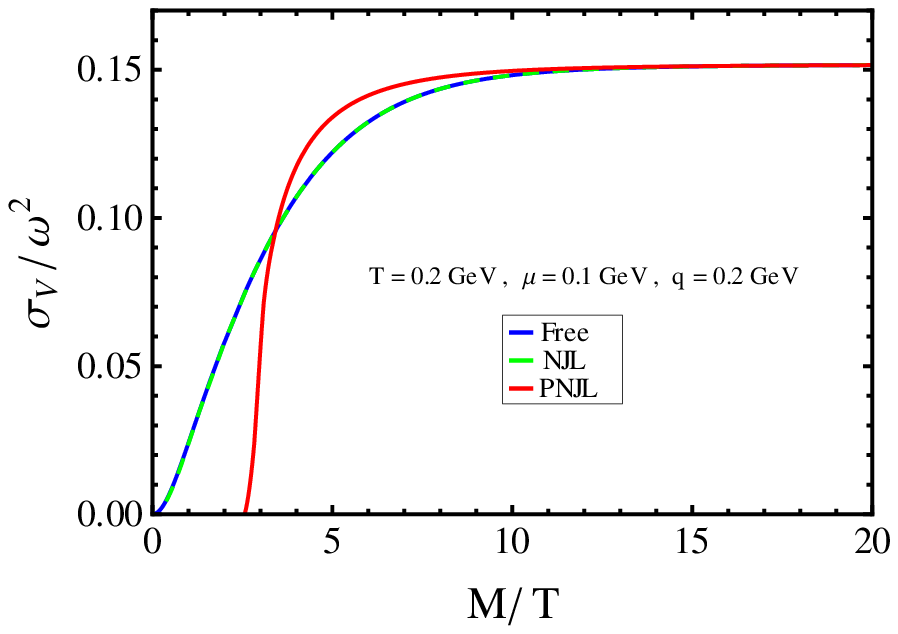}
\label{fg.free_mu1_t2}}
\subfigure[]
{\includegraphics[scale=0.8]{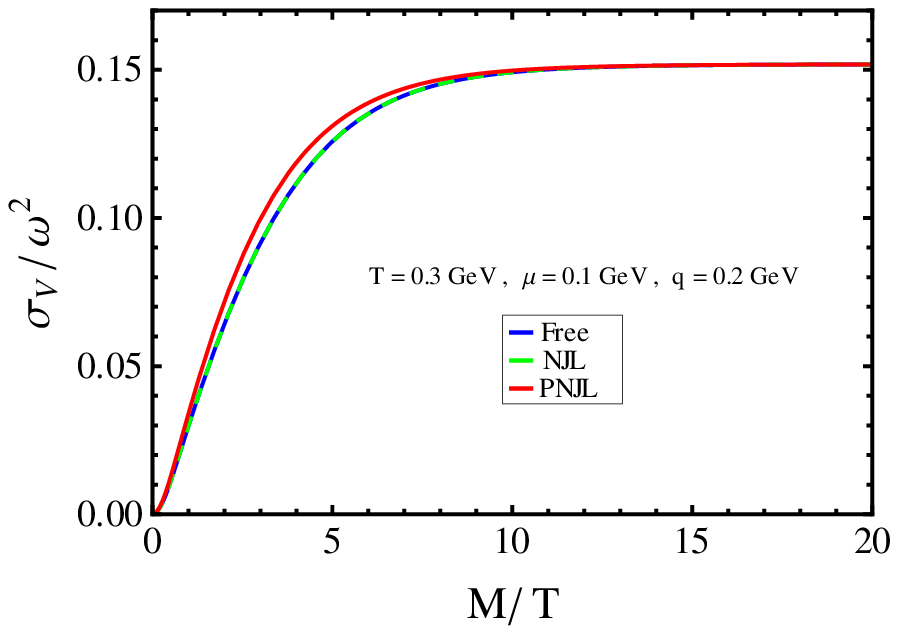}
\label{fg.free_mu1_t3}}
\caption{Scaled vector spectral function $\sigma_V/\omega^2$ as a function of scaled 
invariant mass, $M/T$, in NJL and PNJL model with external momentum 
$q=200$ MeV, quark chemical potential $\mu=100$ MeV and $G_V/G_S=0$ for (a) $T=200$ MeV and (b) $T=300$ MeV}
\label{fg.spect_gv0_q02_mu0}
\end{figure}
 
\subsubsection{Without vector interaction ($G_V=0$)}

With no vector interaction ($G_V=0$), the spectral function in (\ref{eq.spectral_resum}) 
is solely determined by  the imaginary part of the one loop vector self energies 
$\Pi_{00}(\omega,\vec q)$  and $\Pi_{ii}(\omega,\vec q)$.
Fig.~\ref{fg.spect_gv0_q0_mu0} displays a comparison of vector spectral function
with zero external momentum ($\vec q=0, \ \ M=\omega$) in NJL and PNJL model for $T=200$ MeV and $300$ MeV, 
when there is no vector interaction ($G_V=0$). 
Now, for $T=200$ MeV (left panel) the spectral function in PNJL model has larger
threshold than NJL  model because the quark mass in PNJL model is much
larger than that of NJL one (see Fig.~\ref{fg.fields&mass_vs_T}). Also the 
PNJL spectral function dominates over that of NJL one, because of the presence of
nonperturbative effects due to Polyakov Loop fields $\Phi$ and ${\bar \Phi}$. At
higher values of $T~(=300~\rm {MeV})$ (right panel), the threshold becomes almost same due to the 
reduction of mass effect in PNJL case whereas the nonperturbative effects 
at low $M/T$ still dominate. The reason is the following: at zero external momentum and zero
chemical potential the spectral function is proportional to $[1-2f(E_p)]$ (apart from the mass
dependent prefactor) as can be seen from the second
term of (\ref{eq.impi_ii_q0_2}). In PNJL case the thermal distribution function, $f(E_p)$, 
is more suppressed due to the suppression of color degrees of freedom than NJL at moderate
values of $T$, so the weight factor $[1-2 f(E_p)]$ is larger than NJL case and 
causing an enhancement in the spectral function . All these features also 
persist at non-zero chemical
potential and external momentum as can also be seen from Fig.~\ref{fg.spect_gv0_q02_mu0}.

\begin{figure} [!htb]
\vspace*{-0.0in}
\begin{center}
\includegraphics[scale=1.0]{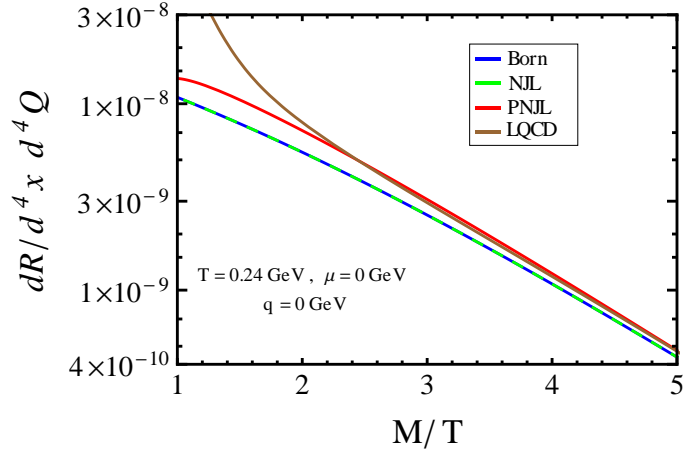}
\end{center}
\caption{Comparison of  dilepton rates  as a function of  $M/T$  
for $T=240$ MeV with external momentum 
$q=0$, quark chemical potential $\mu=0$ and $G_V/G_S=0$. The LQCD rate is from Ref.~\cite{Ding:2010ga}.}
\label{fg.dilep_gv0_q0_mu0}
\end{figure}

At this point it is important to note that for 
$T>250$ MeV the mass in NJL model almost approaches current
quark mass (see Fig.~\ref{fg.massfields_gv0}) and can be considered as a free case 
since there is no vector interaction present ($G_V=0$). Nevertheless, the PNJL case is 
different  because of the presence of the nonperturbative confinement effect
through the Polyakov Loop fields.  The PNJL model can suitably describe a semi-QGP~\cite{Fukushima:2003fw,Fukushima:2003fm,Gale:2014dfa} 
scenario having nonperturbative effect due to the suppression of color degrees of freedom 
compared to NJL vis-a-vis free case above the deconfinement temperature.

The above features of the spectral function  in a semi-QGP with no vector interaction 
will be reflected  in the dilepton rate which is related to the spectral function,  as given 
in (\ref{eq.rel_dilep_spec}). In Fig.~\ref{fg.dilep_gv0_q0_mu0}, the dilepton rate is displayed
as a function of scaled invariant mass $M$ with $T$. As already discussed, at this temperatures 
the quark mass in NJL approaches current quark mass faster than PNJL, 
thus the dilepton rates for Born and NJL cases become almost the same. However, the dilepton rate 
in PNJL model is enhanced than those of Born or NJL case. This in turn suggests that the 
nonperturbative dilepton production rate is higher in a semi-QGP than the Born rate in 
a weakly coupled QGP. The dilepton
rate is also compared with that from LQCD result~\cite{Ding:2010ga} within a quenched approximation.
It is found to agree well for $M/T\ge 2$, below which it differs from LQCD rate. 
We try to understand this as follows:  the spectral function in LQCD 
is extracted using MEM from Euclidean vector correlation function by inverting (\ref{eq.correlation_total}),
which requires an ansatz for the spectral function. Using a free field spectral function as an ansatz, 
the spectral function in a  quenched approximation of QCD was obtained earlier~\cite{Karsch:2001uw}
by inverting (\ref{eq.correlation_total}), which was then approaching zero in the limit $M/T \rightarrow 0$.
So was the first lattice dilepton rate~\cite{Karsch:2001uw} at low $M/T$ whereas it was oscillating around 
the Born rate for $M/T>3$. Now, in a very recent LQCD calculation~\cite{Ding:2010ga} with larger size, 
while extracting the spectral 
function using MEM from Euclidean vector correlation function, an ansatz for the spectral function,
a Briet-Wigner (BW) for low  $M/T $ plus a free field one for $M/T\ge 2$,  has been used. 
The ansatz of BW at  low $M/T$ pushes up the spectral function and so is the recent dilepton 
rate in LQCD below $M/T \le 2$. However, no such ansatz is required in thermal QCD and we can 
directly calculate the spectral function without any uncertainty by virtue of the analytic continuation.

\begin{figure} [!htb]
\vspace*{-0.0in}
\subfigure[]
{\includegraphics[scale=0.8]{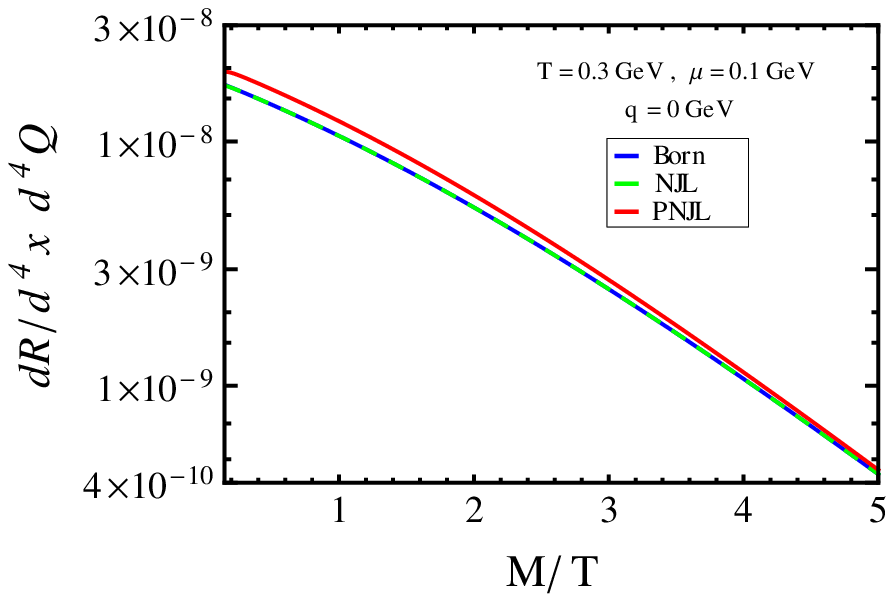}
\label{fg.free_dilepq0mu}}
\subfigure[]
{\includegraphics[scale=0.8]{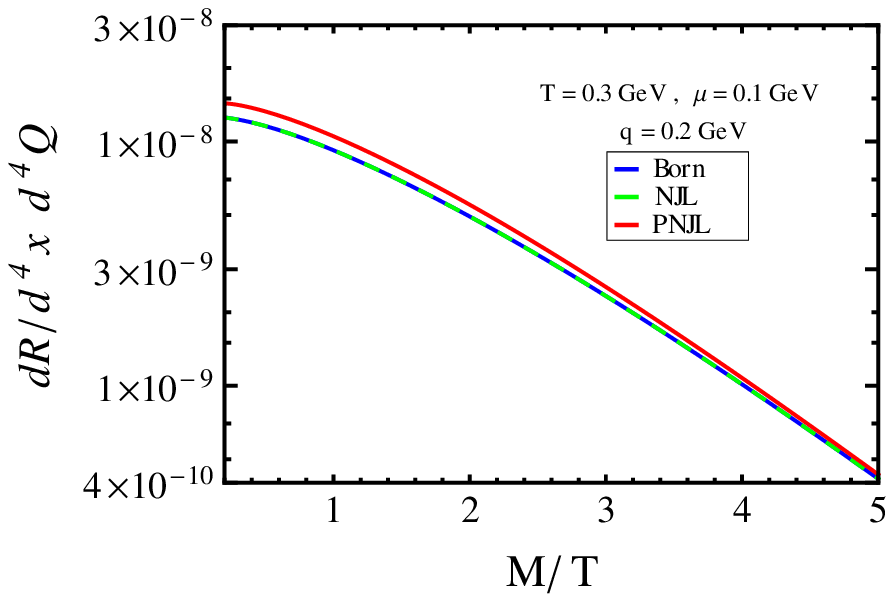}
\label{fg.free_dilepq2mu}}
\caption{Different dilepton rates  as a function of  $M/T$ 
at  $T=300$ MeV, $\mu=100$ MeV and $G_V/G_S=0$ for (a) $q=0$ (b) $q=200$ MeV}
\label{fg.dilep_gv0_mu1}
\end{figure}
In Fig.~\ref{fg.dilep_gv0_mu1} the dilepton rate is also displayed at $T=300$ MeV, 
non-zero chemical potential ($\mu=100$ MeV) and external momentum ($q=0$ and $200$ MeV).
We note that this information could also be indicative for future LQCD 
computation of dilepton rate at  non-zero $\mu$ and $q$.
The similar feature of semi-QGP as found in Fig.~\ref{fg.dilep_gv0_q0_mu0} is 
also seen here but with a quantitative difference especially due to higher $T$,
which could be understood from Fig.~\ref{fg.massfields_gv0}.

\subsubsection{With vector interaction ($G_V\ne 0$)}

\begin{figure} [!htb]
\vspace*{-0.0in}
\subfigure[]
{\includegraphics[scale=0.8]{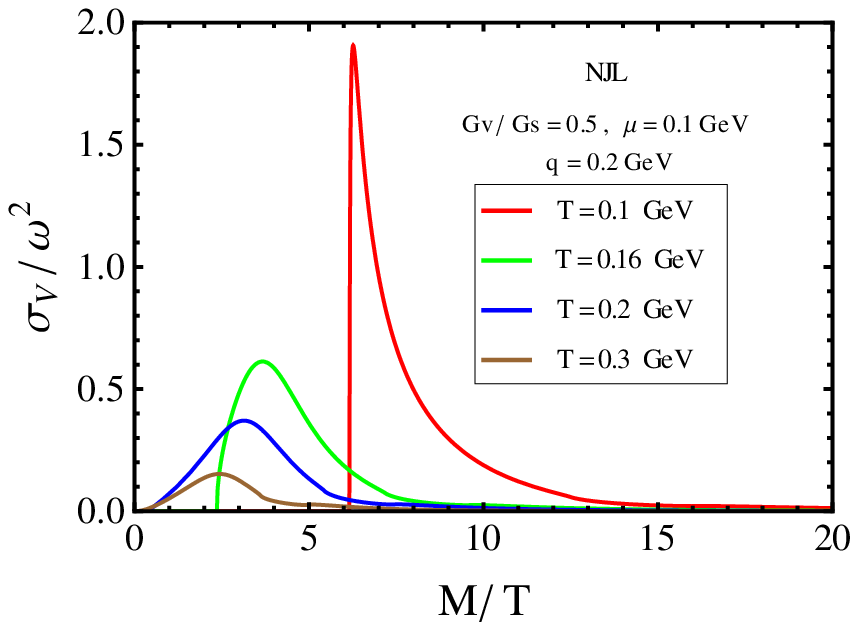}
\label{fg.njl_spect1q2mu1}}
\subfigure[]
{\includegraphics[scale=0.8]{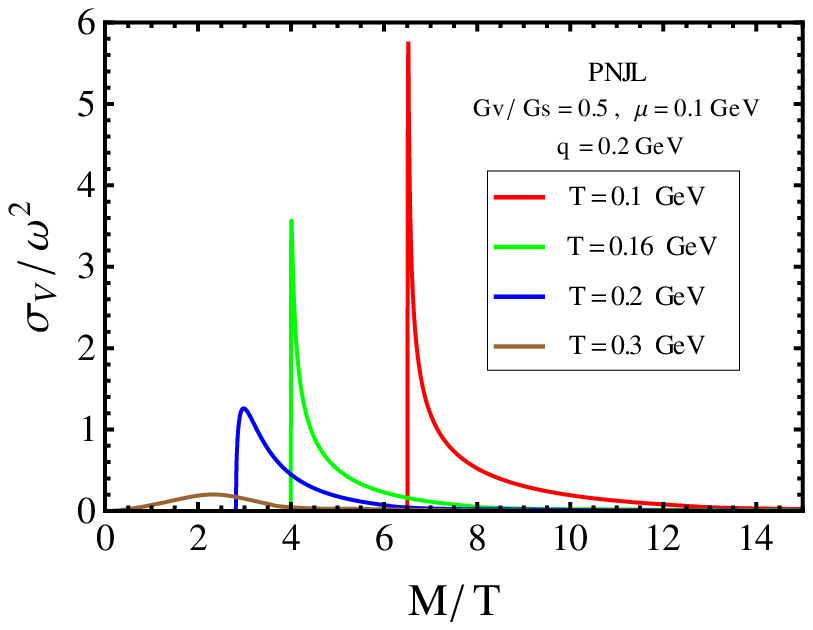}
\label{fg.pnjl_spect1q2mu1}}
\caption{Scaled spectral function  as a function of  $M/T$ in (a) NJL and (b) PNJL model 
for a set of $T$ with 
$\mu=100$ MeV,  $q=200$ MeV and $G_V/G_S=0.5$. Note the difference in $y$-scale.}
\label{fg.spect_njl_gv05}
\end{figure}

\begin{figure} [!htb]
\vspace*{0.1in}
\subfigure[]
{\includegraphics[scale=0.8]{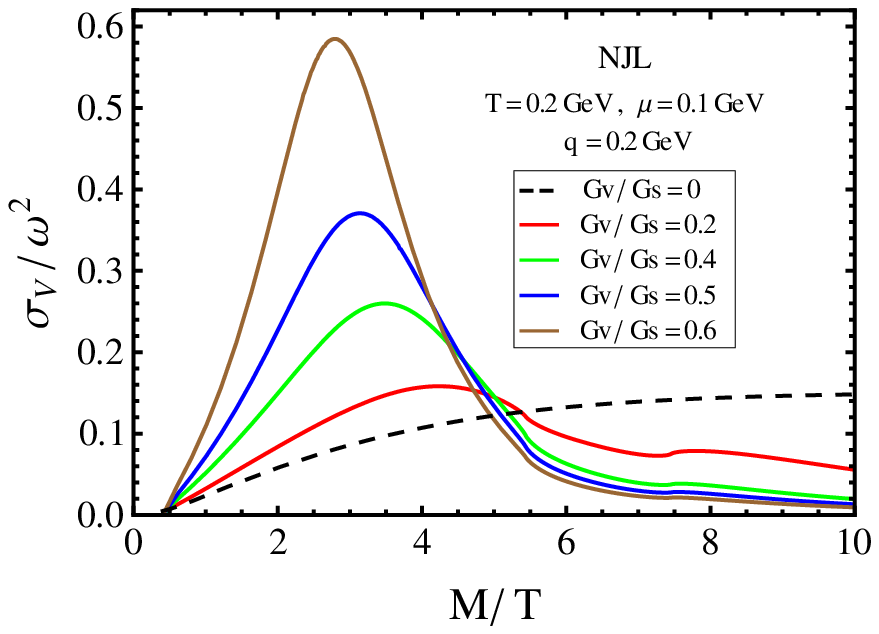}
\label{fg.njl_spect2q2mu1}}
\subfigure[]
{\includegraphics[scale=0.8]{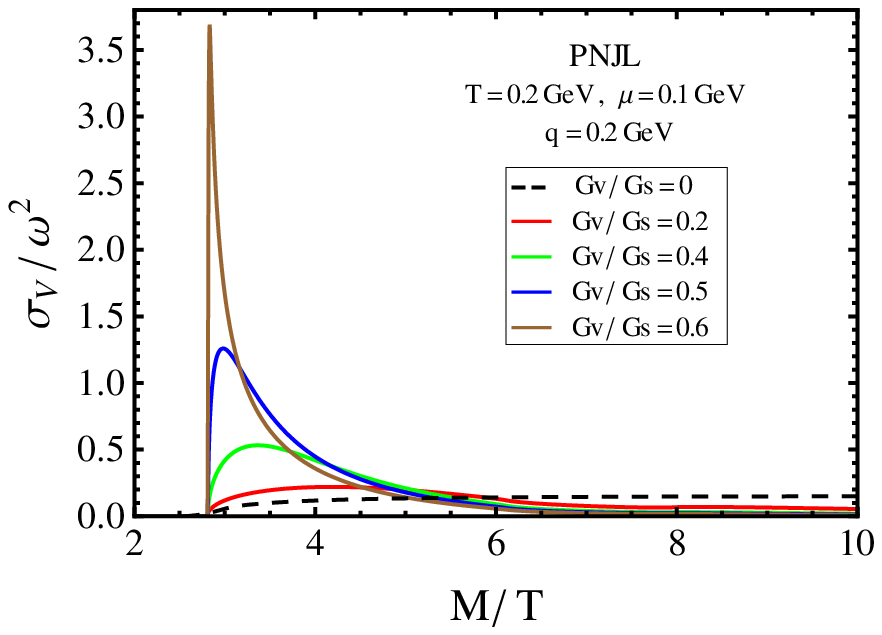}
\label{fg.pnjl_spect2q2mu1}}
\caption{Scaled spectral function  as a function of  $M/T$ 
for (a) NJL  and (b) PNJL model with $T=200$ MeV, 
$\mu=100$ MeV,  $q=200$ MeV and a set of values of $G_V/G_S=0,\, 0.2,\, 0.4, \, 0.5
\, \, {\textrm{and}} \, \, 0.6$.}
\label{fg.spect_t2_q2_mu1}
\end{figure}
\begin{figure} [!htb]
\vspace*{-0.0in}
\subfigure[]
{\includegraphics[scale=0.8]{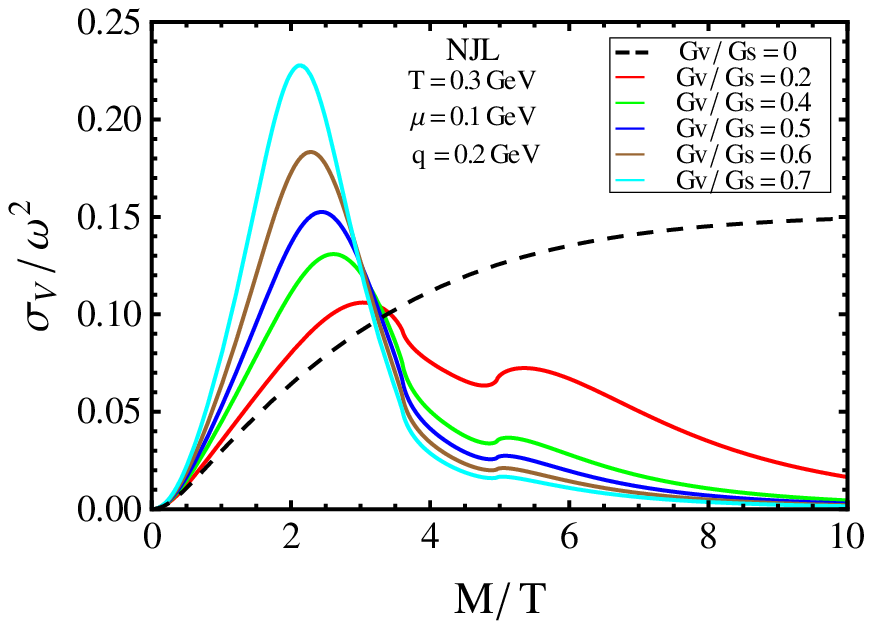}
\label{fg.njl_spect3q2mu1}}
\subfigure[]
{\includegraphics[scale=0.8]{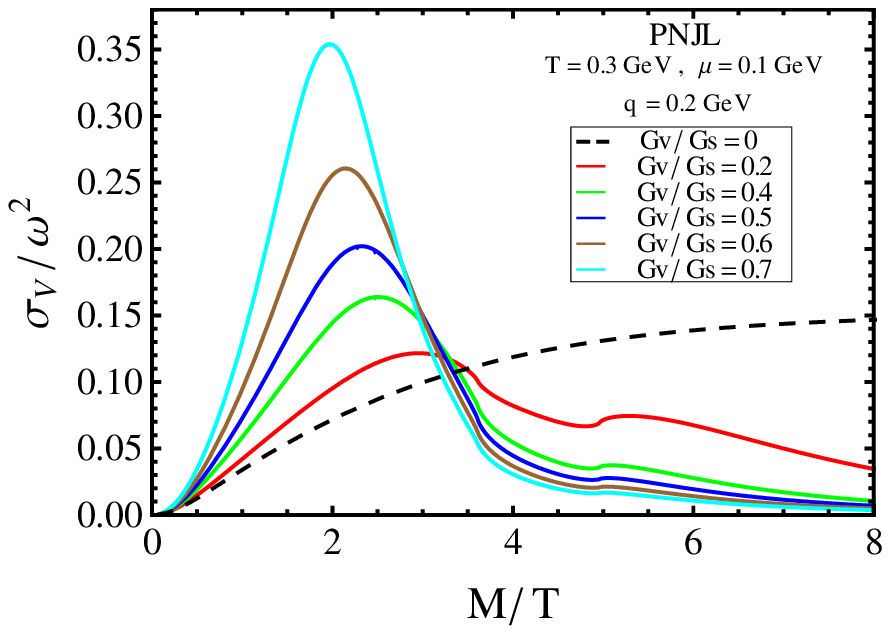}
\label{fg.pnjl_spect3q2mu1}}
\caption{Scaled spectral function  as a function of  $M/T$ 
for (a) NJL  and (b) PNJL model with $T=300$ MeV, 
$\mu=100$ MeV,  $q=200$ MeV and a set of values of $G_V/G_S=0, \ 0.2,\, 0.4, \, 0.5, 
\, 0.6 \, \, {\textrm{and}} \, \, 0.7$.}
\label{fg.spect_t3_q2_mu1}
\end{figure}
\begin{figure} [!htb]
\vspace*{-0.0in}
\subfigure[]
{\includegraphics[scale=0.8]{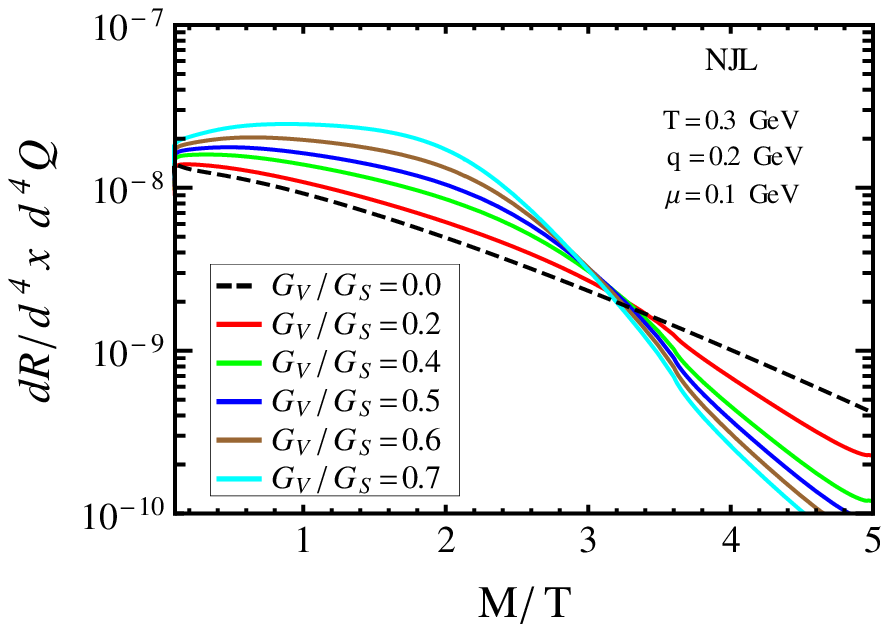}
\label{fg.njl_dilepq0t3}}
\subfigure[]
{\includegraphics[scale=0.8]{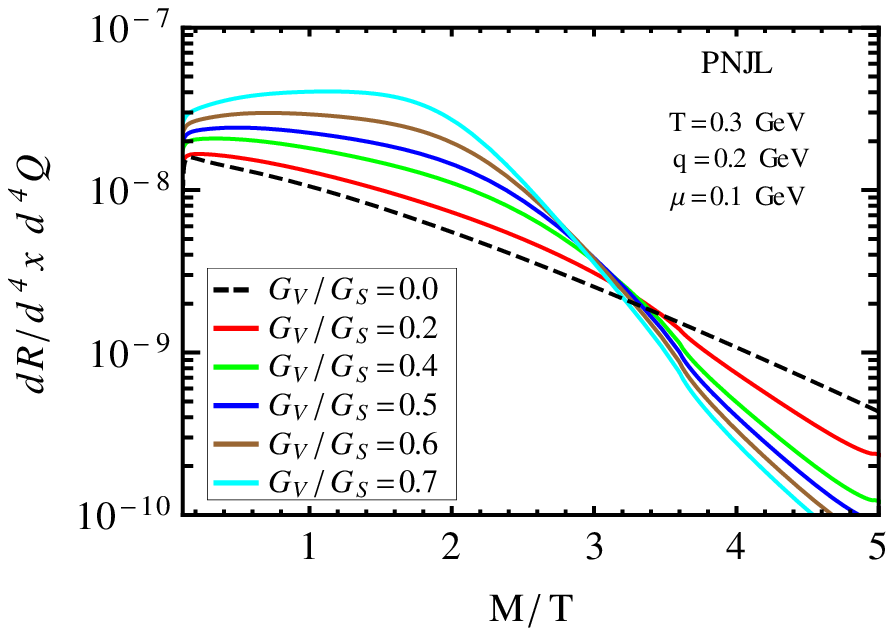}
\label{fg.pnjl_dilepq0t3}}
\caption{Dilepton rates  as a function of  $M/T$  (a) NJL and (b) PNJL model
for a set of values of $G_V/G_S$  with $T=300$ MeV with external momentum 
$q=200$ MeV, quark chemical potential $\mu=100$ MeV.}
\label{fg.dilep_gv_t3}
\end{figure}

In Fig.~\ref{fg.spect_njl_gv05} the spectral function for $G_V/G_S=0.5$ with $q=200$ MeV and 
$\mu=100$ MeV in NJL (left panel) and PNJL (right panel) model is displayed. 
At $T=100 < T_c \sim 160$ MeV~\cite{Islam:2014tea,Bazavov:2011nk,Petreczky:2012ct,Borsanyi:2010bp}
the spectral function  above the respective threshold, $M> 2M_f$, starts with a large value because the
denominator in (\ref{eq.CTprime}) is very small compared to those in (\ref{eq.C00}) and (\ref{eq.CLprime}).
This is due to the two reasons: (i) the first term in the denominator involving real parts of $\Pi$ has zero 
below the threshold that corresponds to a sharp $\delta$-like peak as discussed in (\ref{eq.discrete_spect}), 
thus it also becomes a very small number just above the threshold and 
(ii) the second term involving imaginary parts start building up, which is also very small. However, the increase in 
$T$ causes the spectral function to decrease due to mutual effects of denominator 
(involving both real and imaginary parts of $\Pi$)  and numerator (involving only imaginary parts of $\Pi$). 
On the other hand, with the increase in $T$, the threshold in NJL case reduces quickly as the quark mass 
decreases faster whereas it reduces slowly for PNJL case because the Polyakov Loop fields experience a 
slow variation of the quark mass. So, the  vector meson in NJL model acquires a width earlier
than the PNJL model due to suppression of color degrees of freedom in presence of Polyakov Loop
fields. As seen, for NJL model at $T= T_c \sim 160$ MeV the sharp peak like structure gets 
a substantial width than  PNJL model. This suggests that the vector meson retains its bound properties 
at and above $T_c$ in PNJL  model in presence of $G_V$ along with the nonperturbative effects 
through Polyakov Loop fields.

In Figs.~\ref{fg.spect_t2_q2_mu1} and \ref{fg.spect_t3_q2_mu1} we present the dependence of 
the spectral function on the vector interaction in QGP for a set of values of the coupling 
$G_V$ in NJL (left panel) and PNJL (right panel) model, respectively,
for $T=200$ and  $T=300$ MeV. In both cases the spectral strength increases 
with that of $G_V$. Nevertheless, the strength of the  spectral function in PNJL case at a 
given $T$ and $G_V$ is always stronger than that of NJL model. This suggests that
the presence of the vector interaction further suppresses the color degrees of freedom 
in addition to the Polyakov Loop fields.
The dilepton rates corresponding to $T=300$ MeV are  also displayed in Fig.~\ref{fg.dilep_gv_t3}, which
show an enhancement at low $M/T$ compared to $G_V=0$ case. The enhancement in PNJL case
indicates that more lepton pairs will be produced at low mass ($M/T < 4$) in semi-QGP 
with vector interaction, which would be appropriate for the hot and dense matter likely to be
produced at FAIR energies.

\subsection{Vector Correlation function}

The vector correlation function can be obtained using (\ref{eq.correlation_total}) in the scaled
Euclidean time $\tau T \in [0, 1]$. We note that
the correlation function in the $\tau T$ range is symmetrical around $\tau T=1/2$  
due to the periodicity condition in Euclidean time guaranteed by the kernel 
$\cosh[\omega (2\tau T-1)/2T]/\sinh(\omega/2T)$ in 
(\ref{eq.correlation_total}).


\begin{figure} [!htb]
\vspace*{0in}
\begin{center}
\includegraphics[scale=0.9]{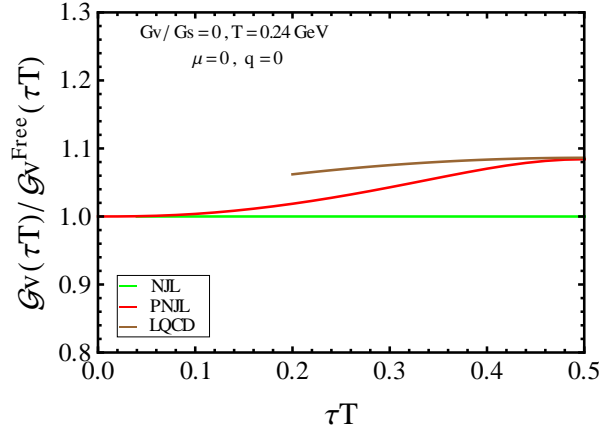}
\end{center}
\caption{Comparison of various scaled Euclidean correlation function with respect to that of 
free field theory, 
${\cal G}_V(\tau T)/{\cal G}_V^{\textrm{free}}(\tau T)$, as a function of the scaled Euclidean time
$\tau T$ for $T=240$ MeV with external momentum 
$q=0$, quark chemical potential $\mu=0$ and $G_V/G_S=0$. The continuum extrapolated 
LQCD result is from Ref.~\cite{Ding:2010ga}.}
\label{fg.corr_gv0_q0_mu0}
\end{figure}

\begin{figure} [!htb]
\begin{center}
\includegraphics[scale=0.9]{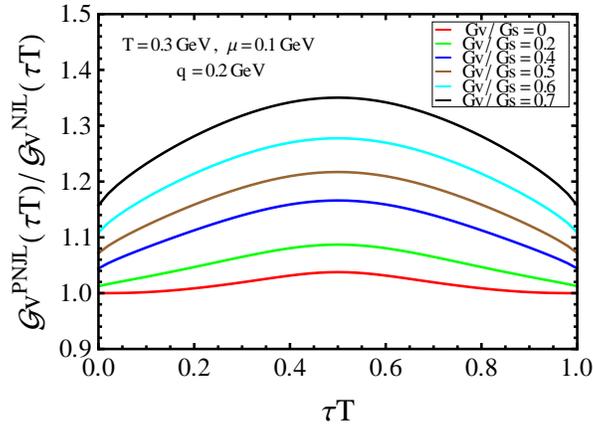}
\end{center}
 \caption{Ratio of Euclidean correlation function in PNJL model to that of NJL model
 as a function of $\tau T$  at $T=300$ MeV,  $\mu=100$ MeV and $q=200$ MeV with a set of values of $G_V/G_S$.}
\label{fg.corr_pnjl_njl}
\end{figure}

In Fig.~\ref{fg.corr_gv0_q0_mu0} a comparison of the ratio of the vector correlation function to that of 
free one is displayed at $T=240$ MeV, $G_V/G_S=0$, $\mu=0$ and $q=0$ for NJL, PNJL and the continuum extrapolated 
LQCD data~\cite{Ding:2010ga} in quenched approximation. It is plotted in the $\tau T$ range $[0,1/2]$ because 
LQCD data are available for the same range. As seen the NJL case becomes
equal to that of free one as there is no effects from background gauge fields. On the other hand
the PNJL result  at $\tau T=0$ and $1$ becomes similar to those of free case because 
$\sigma_V^{\textrm{PNJL}}(\omega)/\sigma_V^{\textrm{free}}(\omega)=1$ as $\omega \rightarrow \infty$.  
Around $\tau T=1/2 $ it deviates maximum from the free case due to the difference in spectral function
at small $\omega $ and thus a nontrivial correlation exists among color charges due to the presence 
of the nonperturbative Polyakov Loop fields. This features are consistent with those in
the spectral function in Fig.~\ref{fg.spect_gv0_q0_mu0} and the dilepton rate in Fig.~\ref{fg.dilep_gv0_q0_mu0}. 
In contrary, the correlation function around $\tau T=1/2$ agrees better with that of LQCD in 
quenched approximation~\cite{Ding:2010ga}.

In Fig.~\ref{fg.corr_pnjl_njl}, we display the effect of the vector interaction $G_V$ in
addition to the presence of the Polyakov Loop fields in QGP through the ratio of the correlation 
function in PNJL model to that of NJL model. Here we have displayed the result in the full range
of the scaled Euclidean time $\tau T \in [0,1]$. As discussed the ratio is symmetric around
$\tau T=1/2$ and always stay above unity. The ratio increases with the increase of the strength of the vector interaction.
This is due to the fact that PNJL correlation function 
is always larger than that of the NJL case since   
$\sigma_V^{\textrm{PNJL}}(\omega,\vec q)/\sigma_V^{\textrm{NJL}}(\omega,\vec q)>1$, and it is even stronger
in particular at small $\omega$ (see Fig.~\ref{fg.spect_t3_q2_mu1}).  
This indicates that the color 
charges maintain a strong correlation  among them due to the presence of both 
Polyakov Loop fields and the vector interaction.  Thus the vector meson retains its 
bound properties in the deconfined phase.

\subsection{Quark number susceptibility and temporal correlation function}

\begin{figure} [!htb]
\subfigure[]
{\includegraphics[scale=0.8]{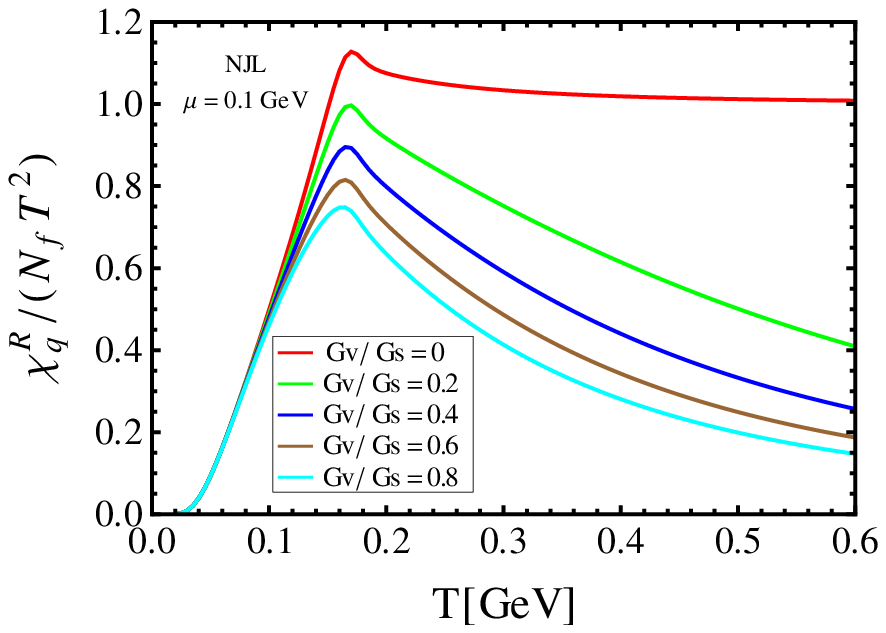}
\label{fg.suscp_njl}}
\subfigure[]
 {\includegraphics[scale=0.8]{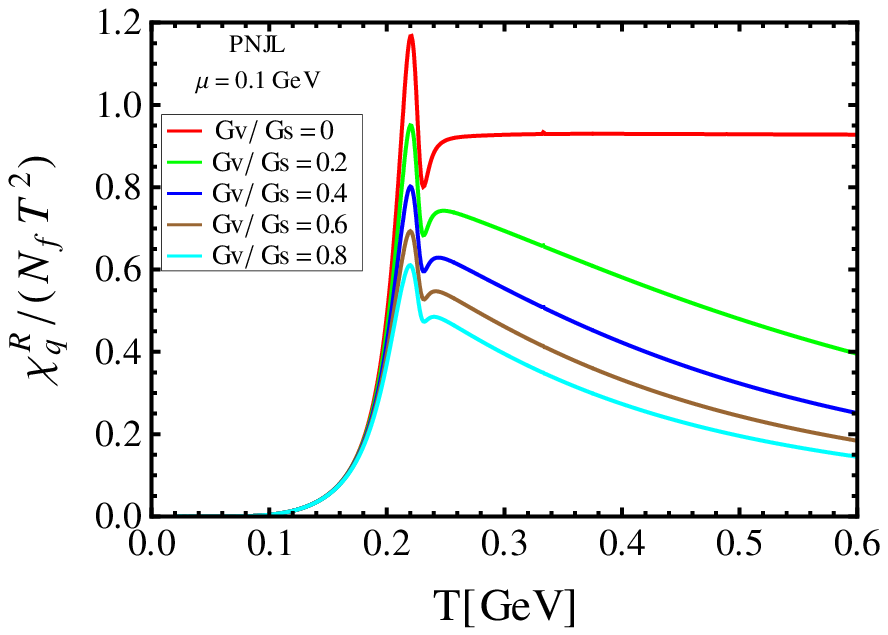}
 \label{fg.suscp_pnjl}}
 \caption{Resummed quark number susceptibility in (a) NJL model
 and (b) PNJL model at non-zero chemical potential for two flavor $(N_f=2)$.}
\label{fg.susceptibility}
\end{figure}
Now, we can calculate quark number susceptibility (QNS) associated with 
the temporal part of the vector spectral function through the
conserved density fluctuation as given in (\ref{eq.resum_suscp}). 
The resummed susceptibility for a set of values of $G_V$ at finite
quark chemical potential ($\mu=100$ MeV) is shown in  
Fig.~\ref{fg.susceptibility}.
For positive vector coupling $G_V$ the denominator of 
(\ref{eq.resum_suscp}) is always greater than unity and
as a result the resummed susceptibility gets suppressed as 
one increases $G_V$. Since positive $G_V$ implies a repulsive
interaction, the compressibility of the system decreases
with increase of $G_V$, hence the susceptibility as seen from
Fig.~\ref{fg.susceptibility} decreases. We note that 
the QNS at finite $\mu$ shows an important feature around
the phase transition temperature than that at $\mu=0$~\cite{Ghosh:2014zra}. 
This is due to the fact that the mean fields  ($X=\sigma, \ \, \Phi$ and $\bar \Phi$)  
which implicitly depend on $\mu$ contributes strongly as the change 
of these fields are most significant around the transition region. 
This feature could be important in the perspective of FAIR scenario
where a hot but very dense matter is expected to be created.
Using (\ref{eq.eucl_chiq}) one can compute the temporal Euclidean 
correlation function associated with the QNS as ${\cal G}^E_{00}(\tau T)/T=-\chi^R_q(T)$, which 
does not depend on $\tau$ but on $T$. This study could also provide useful information to
future LQCD calculation at finite $\mu$.

\section{Conclusion}
\label{sc.conclusion}

In the present work, the behavior of the vector meson 
correlation function and its spectral representation, and various physical 
quantities associated with the spectral representation, in a hot and 
dense environment,  have been studied 
within the effective model framework, {\it viz.} NJL and PNJL model.
PNJL model contains additional nonperturbative information through
Polyakov Loop fields than NJL model. In addition to this nonperturbative 
effect of Polyakov Loop, the repulsive isoscalar-vector interaction
is also considered. The influence of such interaction on the correlator 
and its spectral representation in a hot and dense medium has been obtained 
using ring resummation known as 
Random Phase Approximation. The incorporation of vector interaction 
is important, in particular, for various spectral properties of the system 
at non-zero chemical potential. However, the value of the vector coupling
strength is difficult to fix from the mass scale which is higher than the maximum 
energy scale $\Lambda$ of the effective theory. So, we have made different choices of this 
vector coupling strength to understand qualitatively its effect on the various quantities
we have computed.

In absence of the isoscalar-vector interaction, the static spectral function and the correlation
function in NJL model become quantitatively equivalent to those of free field theory. In case of
PNJL these quantities are different from both free and NJL case because of the presence of 
the nonperturbative Polyakov Loop fields that suppress the color degrees of freedom in 
the deconfined phase just above $T_c$. This suggests that some nontrivial correlation 
exist among the color charges in the deconfined phase. As an important consequence, the nonperturbative 
dilepton production rate is enhanced in the deconfined phase compared to the leading 
order perturbative rate. We note that the Euclidean correlation function and the nonperturbative 
rate with zero chemical potential agree  well with the available LQCD data in quenched approximation. 
We also discussed these quantities in presence of finite chemical potential and external momentum
which could provide useful information if, in future, LQCD computes them at finite chemical potential
and external momentum.

In presence of the isoscalar-vector interaction, appropriate for hot but very dense medium 
likely to be created at FAIR GSI, it is found that the color degrees of freedoms are, further, suppressed
up to a moderate value of the temperature above the critical temperature implying a stronger correlation 
among the color charges in the deconfined phase. The correlation function, spectral function and its spectral
property, {\it e.g.}, the low mass dilepton rate are strongly affected in PNJL case than NJL case. We also note that 
the response to the conserved number density fluctuation at finite chemical potential exhibits an 
interesting characteristic around the phase transition temperature than that at
vanishing chemical potential. This is because the mean fields (Polyakov Loop fields 
and condensates etc.)  depend implicitly on chemical potential and  so their variations
are most significant around the transition region, in particular for PNJL model.
 Finally, some of our results presented in 
this work can be tested  when LQCD computes them, in future, with the inclusion of the dynamical fermions.

 \begin{acknowledgments}
  CAI would like to acknowledge the financial support from the University Grants Commission, India. SM, NH and MGM 
  acknowledge the Department of Atomic Energy, India for the financial support through the
  project name TPAES. SM would like to thank Anirban Lahiri for useful discussions.
  Finally, we would also like to acknowledge very useful discussions and comments
  we had with Purnendu Chakraborty and Aritra Bandyopadhyay during this work.
 \end{acknowledgments}


\begin{thebibliography}{99}

\bibitem{Muller:1983ed} 
  B.~Muller,
  ``The Physics Of The Quark - Gluon Plasma,''
  Lect.\ Notes Phys.\  {\bf 225}, 1 (1985).
  
\bibitem{Heinz:2000bk} 
  U.~W.~Heinz and M.~Jacob,
  ``Evidence for a new state of matter: An Assessment of the results from the CERN lead beam program,''
  nucl-th/0002042.

\bibitem{Arsene:2004fa} 
  I.~Arsene {\it et al.}  [BRAHMS Collaboration],
 ``Quark gluon plasma and color glass condensate at RHIC? The Perspective from the BRAHMS experiment,''
  Nucl.\ Phys.\ A {\bf 757}, 1 (2005)
  [nucl-ex/0410020].
  
\bibitem{Back:2004je} 
  B.~B.~Back, M.~D.~Baker, M.~Ballintijn, D.~S.~Barton, B.~Becker, R.~R.~Betts, A.~A.~Bickley and R.~Bindel {\it et al.},
  ``The PHOBOS perspective on discoveries at RHIC,''
  Nucl.\ Phys.\ A {\bf 757}, 28 (2005)
  [nucl-ex/0410022].
  
 \bibitem{Adams:2005dq} 
  J.~Adams {\it et al.}  [STAR Collaboration],
  ``Experimental and theoretical challenges in the search for the quark gluon plasma: The STAR Collaboration's critical assessment of the evidence from RHIC collisions,''
  Nucl.\ Phys.\ A {\bf 757}, 102 (2005)
  [nucl-ex/0501009].

\bibitem{Adcox:2004mh} 
  K.~Adcox {\it et al.}  [PHENIX Collaboration],
  ``Formation of dense partonic matter in relativistic nucleus-nucleus collisions at RHIC: Experimental evaluation by the PHENIX collaboration,''
  Nucl.\ Phys.\ A {\bf 757}, 184 (2005)
  [nucl-ex/0410003].
  
\bibitem{Carminati:2004fp} 
  F.~Carminati {\it et al.}  [ALICE Collaboration],
  ``ALICE: Physics performance report, volume I,''
  J.\ Phys.\ G {\bf 30}, 1517 (2004).

\bibitem{Alessandro:2006yt} 
  B.~Alessandro {\it et al.}  [ALICE Collaboration],
  ``ALICE: Physics performance report, volume II,''
  J.\ Phys.\ G {\bf 32}, 1295 (2006).
  
\bibitem{Friman:2011zz} 
  B.~Friman, C.~Hohne, J.~Knoll, S.~Leupold, J.~Randrup, R.~Rapp and P.~Senger,
  ``The CBM physics book: Compressed baryonic matter in laboratory experiments,''
  Lect.\ Notes Phys.\  {\bf 814}, 1 (2011).

\bibitem{Adare:2009qk} 
  A.~Adare {\it et al.}  [PHENIX Collaboration],
  ``Detailed measurement of the $e^+ e^-$ pair continuum in $p+p$ and Au+Au collisions at $\sqrt{s_{NN}} = 200$ GeV and implications for direct photon production,''
  Phys.\ Rev.\ C {\bf 81}, 034911 (2010)
  [arXiv:0912.0244 [nucl-ex]].

\bibitem{Adler:2006yt} 
  S.~S.~Adler {\it et al.}  [PHENIX Collaboration],
  ``Measurement of direct photon production in p + p collisions at $\sqrt {s} = 200$-GeV,''
  Phys.\ Rev.\ Lett.\  {\bf 98}, 012002 (2007)
  [hep-ex/0609031].

\bibitem{Adare:2006ti} 
  A.~Adare {\it et al.}  [PHENIX Collaboration],
  ``Scaling properties of azimuthal anisotropy in Au+Au and Cu+Cu collisions at $\sqrt{s_{NN}}$ = 200-GeV,''
  Phys.\ Rev.\ Lett.\  {\bf 98}, 162301 (2007)
  [nucl-ex/0608033].

\bibitem{Adcox:2001jp} 
  K.~Adcox {\it et al.}  [PHENIX Collaboration],
  ``Suppression of hadrons with large transverse momentum in central Au+Au collisions at $\sqrt{s_{NN}}$ = 130-GeV,''
  Phys.\ Rev.\ Lett.\  {\bf 88}, 022301 (2002)
  [nucl-ex/0109003].
  
\bibitem{Adcox:2002au} 
  K.~Adcox {\it et al.}  [PHENIX Collaboration],
  ``Measurement of the Lambda and anti-Lambda particles in Au+Au collisions at $\sqrt{s_{NN}} = 130$-GeV,''
  Phys.\ Rev.\ Lett.\  {\bf 89}, 092302 (2002)
  [nucl-ex/0204007].

\bibitem{Adler:2003kg} 
  S.~S.~Adler {\it et al.}  [PHENIX Collaboration],
  ``Scaling properties of proton and anti-proton production in $\sqrt {s_{NN}}= 200$-GeV Au+Au collisions,''
  Phys.\ Rev.\ Lett.\  {\bf 91}, 172301 (2003)
  [nucl-ex/0305036].

\bibitem{Chujo:2002bi} 
  T.~Chujo [PHENIX Collaboration],
  ``Results on identified hadrons from the PHENIX experiment at RHIC,''
  Nucl.\ Phys.\ A {\bf 715}, 151 (2003)
  [nucl-ex/0209027].
  
\bibitem{Adare:2006nq} 
  A.~Adare {\it et al.}  [PHENIX Collaboration],
  ``Energy Loss and Flow of Heavy Quarks in Au+Au Collisions at $\sqrt {s_{NN}}= 200$-GeV,''
  Phys.\ Rev.\ Lett.\  {\bf 98}, 172301 (2007)
  [nucl-ex/0611018].
  
\bibitem{Abelev:2006db} 
  B.~I.~Abelev {\it et al.}  [STAR Collaboration],
  ``Erratum: Transverse momentum and centrality dependence of high-$p_T$ non-photonic 
  electron suppression in Au+Au collisions at $\sqrt{s_{NN}} = 200$\,GeV,''
  Phys.\ Rev.\ Lett.\  {\bf 98}, 192301 (2007)
  [Erratum-ibid.\  {\bf 106}, 159902 (2011)]
  [nucl-ex/0607012].

\bibitem{Aamodt:2010pb} 
  K.~Aamodt {\it et al.}  [ALICE Collaboration],
  ``Charged-particle multiplicity density at mid-rapidity in central Pb-Pb collisions at $\sqrt{s_{NN}} = 2.76$ TeV,''
  Phys.\ Rev.\ Lett.\  {\bf 105}, 252301 (2010)
  [arXiv:1011.3916 [nucl-ex]].

\bibitem{Aamodt:2010pa} 
  K.~Aamodt {\it et al.}  [ALICE Collaboration],
  ``Elliptic flow of charged particles in Pb-Pb collisions at 2.76 TeV,''
  Phys.\ Rev.\ Lett.\  {\bf 105}, 252302 (2010)
  [arXiv:1011.3914 [nucl-ex]].
  
  \bibitem{Aamodt:2010jd} 
  K.~Aamodt {\it et al.}  [ALICE Collaboration],
  ``Suppression of Charged Particle Production at Large Transverse Momentum in Central Pb--Pb Collisions at $\sqrt{s_{NN}} = 2.76$ TeV,''
  Phys.\ Lett.\ B {\bf 696}, 30 (2011)
  [arXiv:1012.1004 [nucl-ex]].

\bibitem{Aad:2010bu} 
  G.~Aad {\it et al.}  [ATLAS Collaboration],
  ``Observation of a Centrality-Dependent Dijet Asymmetry in Lead-Lead Collisions at $\sqrt{s_{NN}}=2.77$ TeV with the ATLAS Detector at the LHC,''
  Phys.\ Rev.\ Lett.\  {\bf 105}, 252303 (2010)
  [arXiv:1011.6182 [hep-ex]].

\bibitem{Chatrchyan:2011sx} 
  S.~Chatrchyan {\it et al.}  [CMS Collaboration],
  ``Observation and studies of jet quenching in PbPb collisions at nucleon-nucleon center-of-mass energy = 2.76 TeV,''
  Phys.\ Rev.\ C {\bf 84}, 024906 (2011)
  [arXiv:1102.1957 [nucl-ex]].

\bibitem{Forster(book):1975HFBSCF} D. Forster, Hydrodynamics Fluctuation, Broken Symmetry and Correlation Function 
(Benjamin/Cummings, Menlo Park,CA, 1975);

\bibitem{Callen:1951vq} 
  H.~B.~Callen and T.~A.~Welton,
  ``Irreversibility and generalized noise,''
  Phys.\ Rev.\  {\bf 83}, 34 (1951).
  
\bibitem{Kubo:1957mj} 
  R.~Kubo,
  ``Statistical mechanical theory of irreversible processes. 1. General theory and simple applications 
  in magnetic and conduction problems,''
  J.\ Phys.\ Soc.\ Jap.\  {\bf 12}, 570 (1957).

\bibitem{Davidson:1995fq} 
  R.~M.~Davidson and E.~Ruiz Arriola,
  ``Mesonic correlation functions in the NJL model with vector mesons,''
  Phys.\ Lett.\ B {\bf 359}, 273 (1995).

\bibitem{Andronic:2012ut} 
  A.~Andronic, P.~Braun-Munzinger, J.~Stachel and M.~Winn,
  ``Interacting hadron resonance gas meets lattice QCD,''
  Phys.\ Lett.\ B {\bf 718}, 80 (2012)
  [arXiv:1201.0693 [nucl-th]].

\bibitem{Huovinen:2009yb} 
  P.~Huovinen and P.~Petreczky,
  ``QCD Equation of State and Hadron Resonance Gas,''
  Nucl.\ Phys.\ A {\bf 837}, 26 (2010)
  [arXiv:0912.2541 [hep-ph]].

\bibitem{Kapusta_Gale(book):1996FTFTPA} J. I. Kapusta and C. Gale, Finite Temperature Field 
Theory Principle and Applications (Cambridge University Press,
Cambridge, 1996), 2nd ed.

\bibitem{Lebellac(book):1996TFT} M. LeBellac, Thermal Field Theory 
(Cambridge University Press, Cambridge, 1996), 1st ed.

\bibitem{Gale:1990pn} 
  C.~Gale and J.~I.~Kapusta,
  ``Vector dominance model at finite temperature,''
  Nucl.\ Phys.\ B {\bf 357}, 65 (1991).


\bibitem{Mustafa:1999dt} 
  M.~G.~Mustafa, A.~Schafer and M.~H.~Thoma,
  ``Nonperturbative dilepton production from a quark gluon plasma,''
  Phys.\ Rev.\ C {\bf 61}, 024902 (2000)
  [hep-ph/9908461].  
  
  \bibitem{Ghosh:2009bt} 
  S.~Ghosh, S.~Sarkar and S.~Mallik,
  ``Analytic structure of rho meson propagator at finite temperature,''
  Eur.\ Phys.\ J.\ C {\bf 70}, 251 (2010)
  [arXiv:0911.3504 [hep-ph]].

\bibitem{Ding:2010ga} 
  H.-T.~Ding, A.~Francis, O.~Kaczmarek, F.~Karsch, E.~Laermann and W.~Soeldner,
  ``Thermal dilepton rate and electrical conductivity: An analysis of vector current correlation functions in quenched lattice QCD,''
  Phys.\ Rev.\ D {\bf 83}, 034504 (2011)
  [arXiv:1012.4963 [hep-lat]].

\bibitem{Kaczmarek:2011ht} 
  O.~Kaczmarek and A.~Francis,
  ``Electrical conductivity and thermal dilepton rate from quenched lattice QCD,''
  J.\ Phys.\ G {\bf 38}, 124178 (2011)
  [arXiv:1109.4054 [hep-lat]].
  
\bibitem{Francis:2011bt} 
  A.~Francis and O.~Kaczmarek,
  ``On the temperature dependence of the electrical conductivity in hot quenched lattice QCD,''
  Prog.\ Part.\ Nucl.\ Phys.\  {\bf 67}, 212 (2012)
  [arXiv:1112.4802 [hep-lat]].
  
\bibitem{Datta:2003ww} 
  S.~Datta, F.~Karsch, P.~Petreczky and I.~Wetzorke,
  ``Behavior of charmonium systems after deconfinement,''
  Phys.\ Rev.\ D {\bf 69}, 094507 (2004)
  [hep-lat/0312037].
  
\bibitem{Aarts:2007wj} 
  G.~Aarts, C.~Allton, J.~Foley, S.~Hands and S.~Kim,
  ``Spectral functions at small energies and the electrical conductivity in hot, quenched lattice QCD,''
  Phys.\ Rev.\ Lett.\  {\bf 99}, 022002 (2007)
  [hep-lat/0703008 [HEP-LAT]].

\bibitem{Aarts:2002cc} 
  G.~Aarts and J.~M.~Martinez Resco,
  ``Transport coefficients, spectral functions and the lattice,''
  JHEP {\bf 0204}, 053 (2002)
  [hep-ph/0203177].

\bibitem{Aarts:2002tn} 
  G.~Aarts and J.~M.~Martinez Resco,
  ``Ward identity and electrical conductivity in hot QED,''
  JHEP {\bf 0211}, 022 (2002)
  [hep-ph/0209048].
  
\bibitem{Amato:2013naa} 
  A.~Amato, G.~Aarts, C.~Allton, P.~Giudice, S.~Hands and J.~I.~Skullerud,
  ``Electrical conductivity of the quark-gluon plasma across the deconfinement transition,''
  Phys.\ Rev.\ Lett.\  {\bf 111}, 172001 (2013)
  [arXiv:1307.6763 [hep-lat]].
  
\bibitem{Karsch:2001uw} 
  F.~Karsch, E.~Laermann, P.~Petreczky, S.~Stickan and I.~Wetzorke,
  ``A Lattice calculation of thermal dilepton rates,''
  Phys.\ Lett.\ B {\bf 530}, 147 (2002)
  [hep-lat/0110208].

\bibitem{Aarts:2005hg} 
  G.~Aarts and J.~M.~Martinez Resco,
  ``Continuum and lattice meson spectral functions at nonzero momentum and high temperature,''
  Nucl.\ Phys.\ B {\bf 726}, 93 (2005)
  [hep-lat/0507004].

\bibitem{Karsch:2000gi} 
  F.~Karsch, M.~G.~Mustafa and M.~H.~Thoma,
  ``Finite temperature meson correlation functions in HTL approximation,''
  Phys.\ Lett.\ B {\bf 497}, 249 (2001)
  [hep-ph/0007093].
  
\bibitem{Braaten:1990wp} 
  E.~Braaten, R.~D.~Pisarski and T.~C.~Yuan,
  ``Production of Soft Dileptons in the Quark - Gluon Plasma,''
  Phys.\ Rev.\ Lett.\  {\bf 64}, 2242 (1990).

\bibitem{Greiner:2010zg} 
  C.~Greiner, N.~Haque, M.~G.~Mustafa and M.~H.~Thoma,
  ``Low Mass Dilepton Rate from the Deconfined Phase,''
  Phys.\ Rev.\ C {\bf 83}, 014908 (2011)
  [arXiv:1010.2169 [hep-ph]].

\bibitem{Chakraborty:2003uw} 
  P.~Chakraborty, M.~G.~Mustafa and M.~H.~Thoma,
  ``Quark number susceptibility, thermodynamic sum rule, and hard thermal loop approximation,''
  Phys.\ Rev.\ D {\bf 68}, 085012 (2003)
  [hep-ph/0303009].

\bibitem{Chakraborty:2001kx} 
  P.~Chakraborty, M.~G.~Mustafa and M.~H.~Thoma,
  ``Quark number susceptibility in hard thermal loop approximation,''
  Eur.\ Phys.\ J.\ C {\bf 23}, 591 (2002)
  [hep-ph/0111022].

\bibitem{Chakraborty:2002yt} 
  P.~Chakraborty, M.~G.~Mustafa and M.~H.~Thoma,
  ``Chiral susceptibility in hard thermal loop approximation,''
  Phys.\ Rev.\ D {\bf 67}, 114004 (2003)
  [hep-ph/0210159].
  
\bibitem{Laine:2011xm} 
  M.~Laine, A.~Vuorinen and Y.~Zhu,
 ``Next-to-leading order thermal spectral functions in the perturbative domain,''
  JHEP {\bf 1109}, 084 (2011)
  [arXiv:1108.1259 [hep-ph]].

\bibitem{Czerski:2008zz} 
  P.~Czerski,
  ``HTL meson correlation functions at finite momentum and chemical potential,''
  Nucl.\ Phys.\ A {\bf 807}, 11 (2008). 
  
\bibitem{Czerski:2006ct} 
  P.~Czerski,
  ``Pseudoscalar Meson Temporal Correlation Function in HTL approach,''
  Int.\ J.\ Mod.\ Phys.\ A {\bf 22}, 672 (2007)
  [hep-ph/0608134].
\bibitem{Alberico:2006wc} 
  W.~M.~Alberico, A.~Beraudo, P.~Czerski and A.~Molinari,
  ``Finite momentum meson correlation functions in a QCD plasma,''
  Nucl.\ Phys.\ A {\bf 775}, 188 (2006)
  [hep-ph/0605060].
  
\bibitem{Alberico:2004we} 
  W.~M.~Alberico, A.~Beraudo and A.~Molinari,
  ``Meson correlation functions in a QCD plasma,''
  Nucl.\ Phys.\ A {\bf 750}, 359 (2005)
  [hep-ph/0411346].

\bibitem{Arnold:2000dr} 
  P.~B.~Arnold, G.~D.~Moore and L.~G.~Yaffe,
  ``Transport coefficients in high temperature gauge theories. 1. Leading log results,''
  JHEP {\bf 0011}, 001 (2000)
  [hep-ph/0010177].
  
\bibitem{Arnold:2003zc} 
  P.~B.~Arnold, G.~D.~Moore and L.~G.~Yaffe,
  ``Transport coefficients in high temperature gauge theories. 2. Beyond leading log,''
  JHEP {\bf 0305}, 051 (2003)
  [hep-ph/0302165].

\bibitem{Burnier:2012ze} 
  Y.~Burnier and M.~Laine,
  ``Massive vector current correlator in thermal QCD,''
  JHEP {\bf 1211}, 086 (2012)
  [arXiv:1210.1064 [hep-ph]].

\bibitem{Laine:2003bd} 
  M.~Laine and M.~Vepsalainen,
  ``Mesonic correlation lengths in high temperature QCD,''
  JHEP {\bf 0402}, 004 (2004)
  [hep-ph/0311268].

\bibitem{Hansson:1991kb} 
  T.~H.~Hansson and I.~Zahed,
  ``Hadronic correlators in hot QCD,''
  Nucl.\ Phys.\ B {\bf 374}, 277 (1992).

 \bibitem{Haque:2013sja} 
  N.~Haque, J.~O.~Andersen, M.~G.~Mustafa, M.~Strickland and N.~Su,
  ``Three-loop HTLpt Pressure and Susceptibilities at Finite Temperature and Density,''
  Phys.\ Rev.\ D {\bf 89}, 061701 (2014)
  [arXiv:1309.3968 [hep-ph]]. 
  
  \bibitem{Haque:2014rua} 
  N.~Haque, A.~Bandyopadhyay, J.~O.~Andersen, M.~G.~Mustafa, M.~Strickland and N.~Su,
  ``Three-loop HTLpt thermodynamics at finite temperature and chemical potential,''
  JHEP {\bf 1405}, 027 (2014)
  [arXiv:1402.6907 [hep-ph]].
  
  \bibitem{Haque:2013qta} 
  N.~Haque, M.~G.~Mustafa and M.~Strickland,
  ``Quark Number Susceptibilities from Two-Loop Hard Thermal Loop Perturbation Theory,''
  JHEP {\bf 1307}, 184 (2013)
  [arXiv:1302.3228 [hep-ph]].

\bibitem{Haque:2012my} 
  N.~Haque, M.~G.~Mustafa and M.~Strickland,
  Phys.\ Rev.\ D {\bf 87}, no. 10, 105007 (2013)
  [arXiv:1212.1797 [hep-ph]].
  

\bibitem{He:2003jza} 
  B.~He, H.~Li, C.~M.~Shakin and Q.~Sun,
  ``Comparison of temperature dependent hadronic current correlation functions calculated in lattice simulations of QCD and with a chiral Lagrangian model,''
  Phys.\ Rev.\ C {\bf 67}, 065203 (2003)
  [hep-ph/0302105].


\bibitem{He:2002pv} 
  B.~He, H.~Li, C.~M.~Shakin and Q.~Sun,
  ``Calculation of the pseudoscalar isoscalar hadronic current correlation functions of the quark gluon plasma,''
  Phys.\ Rev.\ D {\bf 67}, 014022 (2003)
  [hep-ph/0210387].

\bibitem{He:2002ii} 
  B.~He, H.~Li, C.~M.~Shakin and Q.~Sun,
  ``Calculation of temperature dependent hadronic correlation functions of pseudoscalar and vector currents,''
  Phys.\ Rev.\ D {\bf 67}, 114012 (2003)
  [hep-ph/0212345].
  
\bibitem{Kunihiro:1991qu} 
  T.~Kunihiro,
  ``Quark number susceptibility and fluctuations in the vector channel at high temperatures,''
  Phys.\ Lett.\ B {\bf 271}, 395 (1991).

\bibitem{Oertel:2000jp} 
  M.~Oertel, M.~Buballa and J.~Wambach,
  ``Meson loop effects in the NJL model at zero and nonzero temperature,''
  Phys.\ Atom.\ Nucl.\  {\bf 64}, 698 (2001)
  [Yad.\ Fiz.\  {\bf 64}, 757 (2001)]
  [hep-ph/0008131].

\bibitem{Fukushima:2003fw} 
  K.~Fukushima,
  ``Chiral effective model with the Polyakov loop,''
  Phys.\ Lett.\ B {\bf 591}, 277 (2004)
  [hep-ph/0310121].
  
\bibitem{Fukushima:2003fm} 
  K.~Fukushima,
  ``Relation between the Polyakov loop and the chiral order parameter at strong coupling,''
  Phys.\ Rev.\ D {\bf 68}, 045004 (2003)
  [hep-ph/0303225].


\bibitem{Pisarski:2000eq} 
  R.~D.~Pisarski,
  ``Quark gluon plasma as a condensate of SU(3) Wilson lines,''
  Phys.\ Rev.\ D {\bf 62}, 111501 (2000)
  [hep-ph/0006205]. 

\bibitem{Dumitru:2001xa} 
  A.~Dumitru and R.~D.~Pisarski,
  ``Degrees of freedom and the deconfining phase transition,''
  Phys.\ Lett.\ B {\bf 525}, 95 (2002)
  [hep-ph/0106176].

\bibitem{Dumitru:2002cf} 
  A.~Dumitru and R.~D.~Pisarski,
  ``Two point functions for SU(3) Polyakov loops near T(c),''
  Phys.\ Rev.\ D {\bf 66}, 096003 (2002)
  [hep-ph/0204223].

\bibitem{Dumitru:2003hp} 
  A.~Dumitru, Y.~Hatta, J.~Lenaghan, K.~Orginos and R.~D.~Pisarski,
  ``Deconfining phase transition as a matrix model of renormalized Polyakov loops,''
  Phys.\ Rev.\ D {\bf 70}, 034511 (2004)
  [hep-th/0311223].
  
\bibitem{Gross:1980br} 
  D.~J.~Gross, R.~D.~Pisarski and L.~G.~Yaffe,
  ``QCD and Instantons at Finite Temperature,''
  Rev.\ Mod.\ Phys.\  {\bf 53}, 43 (1981).

\bibitem{Gocksch:1993iy} 
  A.~Gocksch and R.~D.~Pisarski,
  ``Partition function for the eigenvalues of the Wilson line,''
  Nucl.\ Phys.\ B {\bf 402}, 657 (1993)
  [hep-ph/9302233].

\bibitem{Weiss:1980rj} 
  N.~Weiss,
  ``The Effective Potential for the Order Parameter of Gauge Theories at Finite Temperature,''
  Phys.\ Rev.\ D {\bf 24}, 475 (1981).

\bibitem{Weiss:1981ev} 
  N.~Weiss,
  ``The Wilson Line in Finite Temperature Gauge Theories,''
  Phys.\ Rev.\ D {\bf 25}, 2667 (1982).
  
\bibitem{Ghosh:2006qh} 
  S.~K.~Ghosh, T.~K.~Mukherjee, M.~G.~Mustafa and R.~Ray,
  ``Susceptibilities and speed of sound from PNJL model,''
  Phys.\ Rev.\ D {\bf 73}, 114007 (2006)
  [hep-ph/0603050].

\bibitem{Mukherjee:2006hq} 
  S.~Mukherjee, M.~G.~Mustafa and R.~Ray,
  ``Thermodynamics of the PNJL model with nonzero baryon and isospin chemical potentials,''
  Phys.\ Rev.\ D {\bf 75}, 094015 (2007)
  [hep-ph/0609249].


\bibitem{Ghosh:2007wy} 
  S.~K.~Ghosh, T.~K.~Mukherjee, M.~G.~Mustafa and R.~Ray,
  ``PNJL model with a Van der Monde term,''
  Phys.\ Rev.\ D {\bf 77}, 094024 (2008)
  [arXiv:0710.2790 [hep-ph]].


\bibitem{Ratti:2005jh} 
  C.~Ratti, M.~A.~Thaler and W.~Weise,
  ``Phases of QCD: Lattice thermodynamics and a field theoretical model,''
  Phys.\ Rev.\ D {\bf 73}, 014019 (2006)
  [hep-ph/0506234].

\bibitem{Ghosh:2014zra} 
  S.~K.~Ghosh, A.~Lahiri, S.~Majumder, M.~G.~Mustafa, S.~Raha and R.~Ray,
  ``Quark Number Susceptibility : Revisited with Fluctuation-Dissipation Theorem in mean field theories,''
  Phys.\ Rev.\ D {\bf 90}, 054030 (2014)
  [arXiv:1407.7203 [hep-ph]].  


\bibitem{Hansen:2006ee} 
  H.~Hansen, W.~M.~Alberico, A.~Beraudo, A.~Molinari, M.~Nardi and C.~Ratti,
  ``Mesonic correlation functions at finite temperature and density in the Nambu-Jona-Lasinio model with a Polyakov loop,''
  Phys.\ Rev.\ D {\bf 75}, 065004 (2007)
  [hep-ph/0609116].

  \bibitem{Deb:2009ng} 
  P.~Deb, A.~Bhattacharyya, S.~Datta and S.~K.~Ghosh,
  ``Mesonic Excitations of QGP: Study with an Effective Model,''
  Phys.\ Rev.\ C {\bf 79}, 055208 (2009)
  [arXiv:0901.1992 [nucl-th]].
  
  
\bibitem{Carignano:2010ac} 
  S.~Carignano, D.~Nickel and M.~Buballa,
  ``Influence of vector interaction and Polyakov loop dynamics on inhomogeneous chiral symmetry breaking phases,''
  Phys.\ Rev.\ D {\bf 82}, 054009 (2010)
  [arXiv:1007.1397 [hep-ph]].

\bibitem{Kashiwa:2006rc} 
  K.~Kashiwa, H.~Kouno, T.~Sakaguchi, M.~Matsuzaki and M.~Yahiro,
  ``Chiral phase transition in an extended NJL model with higher-order multi-quark interactions,''
  Phys.\ Lett.\ B {\bf 647}, 446 (2007)
  [nucl-th/0608078].

\bibitem{Sakai:2008ga} 
  Y.~Sakai, K.~Kashiwa, H.~Kouno, M.~Matsuzaki and M.~Yahiro,
  ``Vector-type four-quark interaction and its impact on QCD phase structure,''
  Phys.\ Rev.\ D {\bf 78}, 076007 (2008)
  [arXiv:0806.4799 [hep-ph]].

\bibitem{Fukushima:2008wg} 
  K.~Fukushima,
  ``Phase diagrams in the three-flavor Nambu-Jona-Lasinio model with the Polyakov loop,''
  Phys.\ Rev.\ D {\bf 77}, 114028 (2008)
  [Erratum-ibid.\ D {\bf 78}, 039902 (2008)]
  [arXiv:0803.3318 [hep-ph]].
  
\bibitem{Fukushima:2008is} 
  K.~Fukushima,
  ``Critical surface in hot and dense QCD with the vector interaction,''
  Phys.\ Rev.\ D {\bf 78}, 114019 (2008)
  [arXiv:0809.3080 [hep-ph]].

\bibitem{Denke:2013jmp} 
  R.Z. Denke, J. C. Macias, M. B. Pinto,
  ``Critical Line Back-Bending Induced either by Finite Nc Corrections or by a Repulsive Vector Channel,''
 Jour. of Mod. Phys {\bf 4}, 1583 (2013).


\bibitem{Jaikumar:2001jq} 
  P.~Jaikumar, R.~Rapp and I.~Zahed,
  ``Photon and dilepton emission rates from high density quark matter,''
  Phys.\ Rev.\ C {\bf 65}, 055205 (2002)
  [hep-ph/0112308].

\bibitem{Haque:2011iz} 
  N.~Haque, M.~G.~Mustafa and M.~H.~Thoma,
  ``Conserved Density Fluctuation and Temporal Correlation Function in HTL Perturbation Theory,''
  Phys.\ Rev.\ D {\bf 84}, 054009 (2011)
  [arXiv:1103.3394 [hep-ph]].

\bibitem{Nakahara:1999vy} 
  Y.~Nakahara, M.~Asakawa and T.~Hatsuda,
  ``Hadronic spectral functions in lattice QCD,''
  Phys.\ Rev.\ D {\bf 60}, 091503 (1999)
  [hep-lat/9905034].

\bibitem{Asakawa:2000tr} 
  M.~Asakawa, T.~Hatsuda and Y.~Nakahara,
  ``Maximum entropy analysis of the spectral functions in lattice QCD,''
  Prog.\ Part.\ Nucl.\ Phys.\  {\bf 46}, 459 (2001)
  [hep-lat/0011040].
  
\bibitem{Wetzorke:2000ez} 
  I.~Wetzorke and F.~Karsch,
  ``Testing MEM with diquark and thermal meson correlation functions,''
  hep-lat/0008008.

\bibitem{Klevansky:1992qe} 
  S.~P.~Klevansky,
  ``The Nambu-Jona-Lasinio model of quantum chromodynamics,''
  Rev.\ Mod.\ Phys.\  {\bf 64}, 649 (1992).

\bibitem{Hatsuda:1994pi} 
  T.~Hatsuda and T.~Kunihiro,
  ``QCD phenomenology based on a chiral effective Lagrangian,''
  Phys.\ Rept.\  {\bf 247}, 221 (1994)
  [hep-ph/9401310].
  
\bibitem{Buballa:2003qv} 
  M.~Buballa,
  ``NJL model analysis of quark matter at large density,''
  Phys.\ Rept.\  {\bf 407}, 205 (2005)
  [hep-ph/0402234].


\bibitem{Islam:2014tea} 
  C.~A.~Islam, R.~Abir, M.~G.~Mustafa, R.~Ray and S.~K.~Ghosh,
  ``The consequences of $SU(3)$ colorsingletness, Polyakov Loop and $Z(3)$ symmetry on a quark–gluon gas,''
  J.\ Phys.\ G {\bf 41}, 025001 (2014).

 
\bibitem{Das(book):FTFT} Ashok Das,
{\it Finite Temperature Field Theory}, World scientific.

\bibitem{Gale:2014dfa} 
  C.~Gale, Y.~Hidaka, S.~Jeon, S.~Lin, J.-F.~Paquet, R.~D.~Pisarski, D.~Satow and V.~V.~Skokov {\it et al.},
  ``Production and Elliptic Flow of Dileptons and Photons in the semi-Quark Gluon Plasma,''
  arXiv:1409.4778 [hep-ph].

\bibitem{Bazavov:2011nk} 
  A.~Bazavov, T.~Bhattacharya, M.~Cheng, C.~DeTar, H.~T.~Ding, S.~Gottlieb, R.~Gupta and P.~Hegde {\it et al.},
  ``The chiral and deconfinement aspects of the QCD transition,''
  Phys.\ Rev.\ D {\bf 85}, 054503 (2012)
  [arXiv:1111.1710 [hep-lat]].

\bibitem{Petreczky:2012ct} 
  P.~Petreczky,
  ``Numerical study of hot strongly interacting matter,''
  J.\ Phys.\ Conf.\ Ser.\  {\bf 402}, 012036 (2012)
  [arXiv:1204.4414 [hep-lat]].

\bibitem{Borsanyi:2010bp} 
  S.~Borsanyi {\it et al.}  [Wuppertal-Budapest Collaboration],
  ``Is there still any $T_c$ mystery in lattice QCD? Results with physical masses in the continuum limit III,''
  JHEP {\bf 1009}, 073 (2010)
  [arXiv:1005.3508 [hep-lat]].
   
\end{thebibliography}
\end{document}